# Enhancing Axial Flow Fan Performance in Air-Cooled Condensers: Tip Vortex Manipulators and Comparative Analysis of Numerical Simulation and Experimental Testing

**Johannes P. Pretorius**[1]
Department of Mechanical and Mechatronic Engineering,
Stellenbosch University,
Cnr Joubert and Banghoek Road,
Stellenbosch 7600,
South Africa
e-mail: jpp@sun.ac.za

**Sybrand J. van der Spuy (Jr.)**
Department of Mechanical and Mechatronic Engineering,
Stellenbosch University,
Cnr Joubert and Banghoek Road,
Stellenbosch 7600,
South Africa
e-mail: 22729089@sun.ac.za

**ABSTRACT**

*Direct dry-cooled power plants typically operate numerous large-diameter axial flow fans in air-cooled condensers (ACCs) to facilitate the condensation of steam in the plant's thermodynamic cycle. Enhanced fan efficiency may, at scale, lead to significant parasitic power reductions and subsequent increases in plant power output. The performance of these fans may be increased by adding blade tip modifications which act to control blade tip leakage flow. Previous studies have shown that manipulating the tip leakage vortex increases the blade-to-air momentum transfer and stabilizes the air flow structures as it traverses the fan blade. This paper takes a step towards the development of improved tip vortex manipulators for an ACC fan. Three-dimensional computational fluid dynamic (CFD) simulations, and experimental testing of*

---

[1] Corresponding author

*the modified and unmodified fan within an ISO test facility is performed. The results are then utilized to assess the accuracy of the simulation model, visualize the manipulated flow structures at the blade tip, and ultimately quantify the effect of the endplate designs on fan performance. Excellent correlation between numerical and test results for the unmodified fan is observed, and the benefit of blade tip modification on fan performance is again confirmed experimentally. The simulation model however fails to predict improvements in fan performance under modification, and simulated tip leakage flows are investigated for clarification. It is hypothesized that deflection of tip endplates under operation account for differences between simulation and experiment, and that fluid-structure interaction analyses could potentially resolve discrepancies.*

**INTRODUCTION**

Globally, water conservation is receiving increased attention. In the context of power generation applications, some constructors are specifying the use of dry-cooling systems despite abundant water resources. In arid and semi-arid countries such as South Africa, such systems are becoming non-negotiable. Within its fleet of baseload power stations, the more modern South African plants employ direct dry-cooled systems, or so-called air-cooled condensers (ACCs). These mechanical draft systems use large axial flow fans to force air across multiple heat-exchanger bundles, to condense steam at the cold end of the steam power cycle.

Some of these ACCs rank among the largest in the world. Together, the Matimba, Majuba, Medupi and Kusile power station ACCs account for 1200 fans of approx. 9-10 m diameter. Each fan motor, operating virtually continuously under baseload conditions, consumes around 180-250 kW, thereby classifying the ACCs as major parasitic power consumers of the power station. Given the fan size and high number of annual operational hours then, small increases in fan efficiency could result in significant power and cost savings at scale.

It is well documented that secondary flows at the blade tip region of turbomachinery play a significant role on their operational efficiency [1-5]. When focusing on axial flow fans, Venter and Kröger [6] show in an early paper that increased tip clearance cause dramatic reductions in fan efficiency. This behaviour is ascribed to turbulent leakage flows across the fan blades' tips.

More recent studies by Corsini et al. [7], Corsini and Sheard [8] and Corsini and Sheard [9] investigate both experimentally and numerically the impact of passive methods to control tip leakage vortices on ducted axial fans for low-noise industrial applications. Variable and constant height endplates are incorporated at blade tips and the effect on fan performance and noise assessed. Findings include reduced noise, total pressure gains and comparable fan efficiencies compared to unmodified fans. The work further proposed an upper limit for the height of an endplate attached to the tip of an axial flow fan blade as being three times that of the maximum thickness of the airfoil profile at the blade tip. The importance of stability in the structure of the tip leakage vortex (TLV) was highlighted and analyzed using normalized helicity contour maps inserted perpendicularly along the blade tip chord.

An experimental investigation by Wilkinson and Van der Spuy [10] was the first to consider the effects of blade tip endplates on ACC fan performance. Their research tests the effects of endplates protruding on both the pressure and suction-side of a scale ACC fan (B2a design). Results show that pressure-side endplate configurations prove most successful, with increases in maximum fan efficiency of approx. 2%, compared to the unmodified fan operating at the same tip clearance.

Numerical work by Ye et al. [11] evaluates the effect of blade tip groove patterns on the performance of a twin stage variable pitch axial fan. Efficiency improvements and noise increases due to the modifications are reported.

Meyer et al. [12] perform a numerical study which investigates the implementation of tip and trailing edge appendages as passive mechanisms to control tip leakage flows. The work considers another scale ACC fan (M design) and confirm the positive effects of these techniques on fan performance. Results indicate that some of the evaluated tip appendages boost fan performance at flow rates lower than design point, while trailing edge modifications improve performance parameters across the stable operating range, including at design point. Further work by the researchers [13] simulate and test the performance of the same fan, modified with an optimized trailing edge appendage to control tip leakage vortex exit trajectory angle. A significant design point efficiency enhancement of 7.7% is observed, with somewhat reduced efficiency at lower flow rates, compared to the unmodified fan.

Based on the endplate concepts of [10] and [12], Pretorius and Erasmus [14] develop refined endplate designs and perform experimental testing of the modified B2a scale fan. Test setup limitations in [14] are overcome in a follow-up study by Pretorius and Strümpfer [15], where a new test fan hub design allows full speed testing of the modified fan, at scaled tip clearances representative of operational large-scale ACC fans (0.4-0.5% of fan casing diameter). Test results show notable increases in fan performance compared to the unmodified fan, with enhancements in fan static efficiency at design and maximum efficiency points reaching 7.5% and 4.4% respectively.

This study continues the work of [10, 14-15] and aims to make progress towards the development of improved tip endplate designs for an ACC fan. The B2a scale fan is again used as datum test fan. Numerical simulation by steady-state periodic 3D CFD of the datum fan is conducted, with validation by experiment.

The fan is then modified by incorporating endplates at the blade tips to manipulate the tip leakage vortex, with the aim of achieving performance enhancements compared to the unmodified fan. Having presented promising test results in previous work, the pressure-side tip endplate design of [14] is used as basis and refined. A new tip endplate design is also conceptualized based on observations from literature and evaluated.

Numerical simulation by steady-state periodic 3D CFD of the modified B2a test fan is attempted for the first time. Corresponding experimental tests of the modified fan are performed, facilitating a direct comparison of results.

The introduction of numerical simulation of the modified fan (with comparative test results) achieves multiple purposes: it evaluates the accuracy of numerically modelling the tip-modified B2a scale fan, it provides insight into how the datum tip leakage flow is manipulated by the proposed endplate designs, and it reveals the potential and limitations of using numerical modelling as basis to develop effective tip endplate designs.

**FAN DESIGN SPECIFICATIONS**

The B2a fan was developed specifically for ACC applications. The design is based on a free vortex swirl distribution, with adjustments in twist angle to allow for a larger fan operating margin, before stall starts occurring [16]. Table 1 presents the fan design parameters and Figure 1 shows the unmodified fan prior to testing, as mounted in the fan test facility at Stellenbosch University.

The B2a fan blades have a NASA LS GA(W)-2 0413 profile at the blade root and a NASA LS GA(W)-2 0409 profile at the blade tip. The blade transitions linearly from one to the other. Figure 2 shows the mounting points, consisting of threaded inserts which are used to attach the endplates to the fan tips with M3 countersunk screws. After an endplate is mounted, the interface between it and the fan blade tip is sealed using rubber putty.

**ENDPLATE DESIGNS**

Two endplates were designed for simulation and test. This section provides an overview of each design. The endplates are laser cut from 1.2 mm AISI 304 stainless steel and are bowed to align with the circumferential curvature at the blade tips.

**Adjusted Erasmus endplate**

The first endplate design is based on the "Endplate 1" design by Pretorius and Erasmus [14]. The development logic has been described in [14] and [15] and therefore won't be repeated here. In essence, the endplate aims to act as an obstruction to blade pressure surface flow migration, weaken the tip leakage vortex strength and decrease mixing losses. The endplate is pseudo-rectangular, with the top and bottom edges of the endplate shaped according to the corresponding upper and lower NASA LS 0409 GA(W)-2 profiles of the blade at the tip radius. In this study, the protruding pressure-side height has been increased (now denoted "Adjusted Erasmus") from 24 mm to 25.8 mm, in an effort to further improve the positive results observed with the endplate to date. Figure 3 shows the CAD design of the endplate with some major dimensions, as well as the actual endplate.

**TLV stabilization endplate**

Analysis by Corsini et al. [8] on an axial fan blade modified by a fixed-height endplate indicated that a major cause for instability in the tip leakage vortex structure is a disproportionality between the rotational and translational character of the flow that passes over a blade. This may lead to tip leakage vortex bursting, where the primary tip vortex becomes unstable and collapses with a significant release of turbulence into the flow. Such phenomena increase losses at the tip and ultimately lowers fan efficiency.

A new tip endplate design is therefore proposed, based on the premise that systematically feeding the tip leakage flow will reduce the size of the tip vortex as well as stabilize it. The endplate aims to channel the air passing over the blade in such a manner that some of the flow on the pressure-side near the blade tip would leak into the tip gap via holes in the endplate. In this way, controlled amounts of air are fed into the tip leakage vortex. As these impinging flows incrementally join the translational flow in the tip gap, they add contributory rotational components, that will hopefully stabilize the tip leakage vortex and avoid tip leakage vortex bursting. Figure 4 shows the new endplate design, denoted "TLV stabilization endplate" hereafter.

**EXPERIMENTAL TEST SETUP AND METHODOLOGY**

**Fan test facility**

This section describes the fan test facility used to conduct the experiments for this study, and on which the simulation domain geometry is based. The large fan test facility at Stellenbosch University is certified as an ISO 5801, Type A (open inlet, open outlet) construction. It is regularly used to test scale models of large-diameter axial-flow fans for industrial air-cooling applications. Figure 5 provides a schematic overview of the facility.

Consider the "side elevation" view of Figure 5. Air enters at the calibrated inlet bell mouth (1), where a pressure transducer measures the static pressure drop. The air then moves through a diffuser and flow control louvres (2). A 7.5 kW auxiliary fan (3), which is equipped with its own flow straighteners, counteracts the pressure losses along the inlet to the facility by acting as a booster to the flow. The air is then diffused up to the flow guide vanes (4) which align the air stream with the facility walls. Wire screens (5) are used for further flow alignment as it enters the plenum (settling) chamber (6). The ambient temperature and the static pressure difference in the plenum chamber are measured, which are used to calculate the density of the air flowing through the fan ($\rho_{plen}$). The test fan is mounted in the outlet bell mouth (7) of the facility, where the flow exhausts to the atmosphere. The test fan is driven by a 10 kW electric motor coupled to a variable speed drive. Here, two more parameters are measured: the fan shaft speed and the fan shaft torque.

Two Endress Hauser PMD75 pressure transducers are employed to measure the static pressure differences to the atmospheric pressure. Ambient conditions are monitored via a mercury barometer and thermometer. Torque measurements are facilitated by an HBM T22 torque transducer, while shaft speed is determined using a proximity sensor. Data acquisition is executed using an HBM PMX system, and subsequent data analysis is performed utilizing CATMAN software. To ensure accuracy, zero measurements are conducted both prior to and following the testing procedure.

**Measurement accuracies**

Measurement accuracies of the facility's test equipment are summarized in Table 2.

**Data processing method**

The test data is processed according to the following section, as per the ISO 5801 standards.

The air mass flow rate through the test facility is obtained from:

$$\dot{m} = \alpha\varepsilon \frac{(\pi d_{bell}^2)}{4} \sqrt{2\, \rho_{amb} \Delta p_{s,bell}} \qquad (1)$$

where $\alpha\varepsilon = 0.9802$ is the calibration constant of the inlet bell mouth, $d_{bell} = 1.008\ m$ is the inlet bell mouth diameter and $\rho_{amb}$ is the ambient density. This density is determined from the average ambient measurements before and after the test, using:

$$\rho_{amb} = \frac{p_{amb}}{R_a T_{amb}} \quad (2)$$

where $R_a = 287.08 \, J/kgK$ the specific gas constant of air. The air density in the plenum chamber is calculated from:

$$\rho_{plen} = \frac{(p_{amb} + \Delta p_{s,plen})}{R_a T_{amb}} \quad (3)$$

The volumetric flow rate through the facility is evaluated as follows:

$$\dot{V} = \frac{\dot{m}}{\rho_{plen}} \quad (4)$$

The velocity in the setting chamber is small, but still considered in the calculation of the dynamic pressure at this point, with:

$$p_{d,plen} = \frac{0.5}{\rho_{plen}} \left(\frac{\dot{m}}{A_{plen}}\right)^2 \quad (5)$$

where $A_{plen} = 13.69 \, m^2$ is the plenum chamber area. As per the ISO 5801 standard, the static pressure difference over the fan may be calculated as the difference between the total pressure in the plenum chamber and atmospheric pressure:

$$\Delta p_{Fs} = p_{amb} - p_{tot,plen} \quad (6)$$

which can also be evaluated as:

$$\Delta p_{Fs} = p_{amb} - (\Delta p_{s,plen} + p_{d,plen}) \quad (7)$$

and reduced to:

$$\Delta p_{Fs} = \Delta p_{s,plen} - p_{d,plen} \quad (8)$$

The fan power is obtained from:

$$P = \left(\frac{2\pi N}{60}\right)\tau \quad (9)$$

The tested volumetric flow rate, fan static pressure difference and power values are scaled to the reference density and rotational speed of the fan using the fan laws, as follows:

$$\dot{V}' = \dot{V}\left(\frac{N'}{N}\right) \quad (10)$$

$$\Delta p'_{Fs} = \Delta p_{Fs}\left(\frac{N'}{N}\right)^2 \left(\frac{\rho'}{\rho_{plen}}\right) \quad (11)$$

$$P' = P\left(\frac{N'}{N}\right)^3 \left(\frac{\rho'}{\rho_{plen}}\right) \quad (12)$$

The fan static efficiency is established from:

$$\eta'_{Fs} = \eta_{Fs} = \frac{\Delta p'_{Fs}\dot{V}'}{P'} \quad (13)$$

The flow coefficient is calculated using ([18]):

$$\phi = \frac{u_z}{u_c} \quad (14)$$

where $u_z$ and $u_c$ represent the axial air velocity through the annulus and the circumferential blade tip velocity respectively. The latter is calculated using:

$$u_c = r_t\left(\frac{2\pi N}{60}\right) \quad (15)$$

where $r_t$ is the tip radius of the fan. The static pressure coefficient is determined from:

$$\psi_{Fs} = \frac{\Delta p'_{Fs}}{0.5\rho' u_c^2} \quad (16)$$

**Experimental repeatability, quality assurance and uncertainty analysis**

The measurement equipment was calibrated before tests were conducted, and each experiment was repeated twice to minimize the random error in the various readings taken. To ensure reliability and a correct testing method, experimental results for the unmodified fan tests were compared between successive tests, as well as to test results from previous work on the same fan configuration. Averages of repeated runs are presented as the final performance characteristics.

The identical uncertainty assessment carried out in [14] remains applicable to the current investigation, given that identical measurements were conducted utilizing identical apparatus within the same test facility. Details concerning this uncertainty evaluation are discussed in [14], thus only a succinct overview of the outcomes will be furnished herein for the sake of comprehensiveness.

A sequential perturbation RSS uncertainty analysis, in accordance with the methodology described in [19], yields the subsequent estimations of uncertainties, as follows:

- A 1.0% margin of uncertainty for the flow coefficient.
- A 2.1% margin of uncertainty for the fan static pressure coefficient.
- A 1.3% margin of uncertainty for the fan power.
- A 2.7% margin of uncertainty for the fan static efficiency.

These uncertainties are graphically represented as error bars in all ensuing performance curves.

**NUMERICAL MODEL**

The numerical model was developed using Ansys Fluent 2022 R1 and is described in the following subsections.

**Computational domain**

To facilitate direct comparison between numerical and experimental results, the scale fan is simulated within the fan test facility, with domain as shown in Figure 6. To minimize computational expense, the symmetric nature of the in-situ scale fan is utilized to model 1/8 sections of the test facility inlet, outlet and rotor geometry. The inlet domain dimensions are based on the facility's plenum chamber, which includes a bell mouth at the inlet to the fan casing. The rotor domain includes a single blade instance. The outlet of the test facility, which is open to atmosphere, is modelled using a section of an extended cylindrical domain. Care is taken to extend the outlet domain far enough to avoid the effects of boundaries on the solution. Previous studies at Stellenbosch University ([18], [20], [21]) have confirmed the accuracy of numerical models that employ similar computational domains.

**Boundary conditions**

Since a 1/8 fan model was used, periodic boundary conditions were implemented on the circumferential faces of the inlet, rotor, and outlet domains. The rotor domain connects to the inlet and outlet domains via moving reference frame interfaces. All walls are modelled as no-slip and stationary relative to the adjacent cell zone, except for the rotor casing, which is assigned a zero absolute rotational velocity to avoid its rotation within the rotating rotor domain.

Table 3 summarizes specific boundary condition values used in the model. The specified mass flow rate (1/8 of full mass flow rate) and rotational speed values are based on the design conditions of the scale fan, while the turbulence intensity specification is based on previous work by [22].

**Computational mesh**

The entire domain was spatially discretized using the unstructured polyhedral mesh option within ANSYS Fluent, with refinement in the regions surrounding the bell mouth inlet and outlet, as well as inflation layers along all wall sections. A fine mesh resolution was also included over the blade surface and hub, and especially in the blade tip region to capture tip leakage flows accurately.

**Mesh sensitivity study, selected resolution, and mesh metrics**

A study was conducted to evaluate the sensitivity of the numerical solution to the mesh resolution. The study considered the effect of overall mesh fineness on the fan static pressure and shaft torque (to evaluate fan power) for the unmodified fan operating at design conditions and tip clearance.

Multiple mesh refinements between 1.05 million and 7.8 million elements were considered, and the parameters of interest for the finer meshes plot as a function of inverse mesh element count (effectively average grid spacing), as shown in Figures 7 and 8. Results indicate numerical uncertainties of 1.4% for both fan static pressure and shaft torque on the finest mesh (7.8 million elements), compared to a linearly extrapolated infinitely fine mesh. This mesh resolution was therefore deemed fine enough to represent an accurate numerical solution, and consequently selected for simulation of the unmodified and modified fan configurations. Minor cell count differences are observed upon inclusion of endplates at the blade tip, with element counts reaching 8.2 million cells for these configurations.

Table 4 presents the element count and major quality metrics for the meshes of the different simulation cases. Regional mesh resolutions of the finest mesh are respectively depicted in Figures 9-12 at the inlet-rotor-outlet domain interfaces, blade surface, trailing edge, and tip region.

**Solution method**

The numerical model is set up as a Reynolds-Averaged Navier-Stokes (RANS) simulation using the realizable $k - \varepsilon$ turbulence model, with enhanced wall treatment. Table 5 provides further details of the solution method and discretization settings employed. The computational setup implements learnings from previous studies on ACC fans that found good correlation between numerical and experimental results ([18], [20], [21]).

Simulations are performed for a number of flow coefficient values that represent the typical operating range of an ACC fan. Solutions are run until monitors for static pressure at the inlet and sum of the torque on the hub and blade stabilize. Any residual periodic variation of these monitored parameters after the solution has converged are averaged over the last 300 iterations.

**RESULTS**

The following section presents the simulation and experimental results for the unmodified and modified test fan.

**Unmodified fan**

Figures 13 and 14 show the experimental results for the unmodified test fan as measured in the fan test facility. The results compare well with those of previous tests performed at Stellenbosch University on the same fan ([10, 14-15]), while also presenting the expected trend of increased fan static pressure and efficiency with decreasing tip clearance. These observations serve as confirmation of the correct test setup and method.

Comparative numerical and experimental results for the unmodified fan, operating with 3 mm tip clearance, are shown in Figures 15-17. The same comparison for operation at 8 mm tip clearance is presented in Figures 18-20. In general, a very good correlation between the simulation and test results are obtained. Simulation predictions are particularly accurate for the fan static pressure coefficient, with some reduced precision for calculated fan power values, resulting in in slight differences in fan static efficiency. At 3 mm tip clearance, the simulated static efficiency is within 1.5% of the test result at the design flow coefficient, while the predicted maximum fan efficiency is 1.9% higher than the experimental value. The corresponding differences at 8 mm tip clearance are 3.0% and 0.4% (lower than test value) respectively.

**Modified fan**

Comparative numerical and experimental results for the modified fan – with endplates attached to the blade tips – are presented in this section. Figures 21-23 present the results for the fan modified with the Adjusted Erasmus endplate, while Figures 24-26 reveal the findings for the fan modified with the TLV stabilization endplate.

As previously noted, ACC fans typically operate at relatively large tip clearances (0.4-0.5% of fan casing diameter), which scale to approx. 8 mm on the test fan. The modified fan is therefore tested and simulated only at this tip clearance, thereby providing a realistic picture of the expected effect of the tip vortex manipulators. Care is taken to ensure that the thickness of the endplate does not reduce the tip clearance to less than 8 mm when the test fan is mounted or simulated.

Figures 21-26 include curves for the unmodified fan test results at 8 mm tip clearance for convenient comparison with modified fan performance.

Experimental results for the modified fan clearly indicate the benefit of tip vortex manipulation using endplates, which correlate with findings of previous work ([14, 15]). Test results, for the fan modified with the Adjusted Erasmus endplate, show a 5.2% increase in fan static efficiency at design flow coefficient, and 2.3% enhancement in maximum fan efficiency, compared to the unmodified fan. Improvements in fan performance, albeit somewhat less prominent, are also observed experimentally for the fan modified with the TLV stabilization endplate. Here, efficiency is boosted by 1.5% at design flow coefficient and 1.5% for maximum fan efficiency.

The numerical model however fails to predict any improvement in fan performance for the modified fan, compared to the unmodified fan. Predictions for fan static pressure coefficient are comparable (with Adjusted Erasmus endplate) or below (with TLV stabilization endplate) the corresponding unmodified fan performance curves. Numerical estimates of fan power are slightly above the experimental values for the unmodified fan at design point, while thereafter following the experimental trend of the power curves for the modified fan towards lower flow rates. The combination of similar fan static pressure and higher fan power observations from simulation, results in reduced fan efficiencies compared to test values for the modified fan.

In an effort to clarify the noted discrepancies between simulated and tested fan performance under tip modification, the next section analyses the simulated flow structures at the blade tip of the fan.

**ANALYSIS OF NUMERICAL RESULTS**

Figures 27-28 present contour plots of static pressure from the unmodified and modified (Adjusted Erasmus endplate) fan simulations, for comparison. The pressure plots for the two modified fan configurations are similar. From these figures, the distinction between the pressure and suction-sides of the fan blades are clearly visible. For the unmodified fan, a large region of low pressure on the suction-side is noted, while the modified fan seems to restrict this low-pressure region to a limited area close to the fan blade surface. A sharp pressure drop in the tip gap, near the leading edge of the unmodified fan is also noted, whereas the endplate appears to mitigate against this for the tip leakage flow of the modified fan.

Contour lines of velocity magnitude are shown in Figures 29-31, for the unmodified and modified fan configurations. These figures indicate that velocities formed in the tip gap are generally an order of magnitude larger compared to the velocities in the rest of the rotor's flow domain. All figures also reveal the formation of tip leakage vortices on the blade suction-side, though differences are clear between the unmodified and modified fans. For the unmodified fan, velocities in the tip gap are high and the resulting tip leakage flow from the pressure-side protrude significantly above the blade suction surface to create a large vortex with oblong shape.

In comparison, the incorporation of the Adjusted Erasmus endplate reduces the tip gap velocities and restricts the formation distance of the vortex such that it remains close to the blade surface and produces a circular shaped vortex. The TLV stabilization endplate creates a similar vortex structure to the one produced by the Adjusted Erasmus endplate. The vortex remains close to the blade suction surface but is slightly larger. The TLV stabilization endplate also produces a small region of high tip gap velocity near the leading edge.

The shape of the vortical structure over the unmodified fan, and the reduction in tip gap velocity with modification, agrees with the findings of [8-9].

Normalized helicity is a parameter that has been used in previous studies ([7-8], [12-13]) to visualize and quantify tip vortices at axial fan blade tips. It is defined as $H_{norm} = (\vec{U} \cdot \vec{v})/(|\vec{U}||\vec{v}|)$, with $\vec{U}$ and $\vec{v}$ denoting the respective absolute velocity and vorticity vectors, and $|\vec{U}|$ and $|\vec{v}|$ representing their respective magnitudes.

Figures 32-34 now depict contours of normalized helicity for the unmodified and modified B2a fan configurations. The contours clearly illustrate two clockwise-rotating tip leakage vortices on the pressure and suction-sides of the blade for all fan configurations. The effect of the two endplate designs on these vortices is very similar. Their inclusion appears to strengthen and enlarge the vortices forming on the pressure-side, compared to the unmodified fan. In line with the analysis for velocity, the plots also clearly show how the endplates form smaller, circular suction-side vortices close to the surface of the blade compared to the larger, irregularly shaped vortices for the unmodified fan.

The above observations agree with some of the findings in literature, where [12] also reports two clockwise-rotating vortical structures, and a strengthening of the pressure-side vortex with inclusion of a constant height pressure-side endplate. The reduction in suction-side vortex size under modification were not found in [12]. The noted phenomena of vortex deviation and vortex direction reversal, as reported by [8] when modifying the blade tip with a constant height pressure-side endplate, are not repeated under similar modification in this work. However, a reduced suction-side vortex closer to the blade surface is consistent with the effects noted in [9] for their "TFmvb" or "multi-vortex-breakdown" endplate.

**DISCUSSION**

The experimental results obtained from the unmodified fan exhibit typical performance characteristics, demonstrating concurrence with previous test findings related to the same fan. Consequently, these results are deemed trustworthy. Test results for the modified fan show improved fan performance characteristics compared to the unmodified fan, which are in turn aligned with results from previous test work ([14-15]) which considered blade tip modifications on the same fan. Numerical and experimental results for the unmodified fan of this study agree very well, which suggests that the simulation model is validated and of sufficient accuracy to provide reliable predictions of modified fan performance. Yet, the simulations of the modified fan do not show any performance benefit compared to the unmodified fan.

An analysis of the simulation data depicts differences between the tip leakage vortex structures of the unmodified and modified fans. Specifically, a decrease in the tip gap velocity, as well as the movement and reduction of the suction-side tip vortex under modification is evident. Some of the noted changes to the tip vortices when modifying the blade tip agrees with observations from literature, while other effects are not previously found. The difference in fan design between the various investigations could play a role on the specific tip effects observed during simulation. Regardless, the potential to manipulate the tip leakage flow and vortical structures by blade tip modification seems clear.

Considering the numerical and experimental results of the two endplate designs evaluated in this investigation, the Adjusted Erasmus endplate outperforms the TLV stabilization endplate. The attempt to stabilize the tip vortex by feeding flow from the pressure-side of the blade into the tip gap therefore seems ineffective, and simulation plots show little difference in the tip leakage flow structures between the two endplates.

In terms of the numerical model's inability to predict fan performance enhancements with tip modification, we can present the following hypothesis. During operation of the specific B2a test fan, some deflection of the fan blades and (potentially) endplates occur. This is inferred from instances where some endplate designs have scraped the fan casing, when testing at low tip clearances. It is therefore hypothesized that this deflection results in an effective tip clearance decrease, with corresponding change to the tip leakage vortex, which improves fan performance compared to the unmodified B2a fan. The current numerical model does not take these operational deflections into consideration, which potentially explains its inability to show increases in fan performance with tip modification. Future studies could consider simulating the performance of a modified fan in its deflected state or using fluid-structure interaction analyses to confirm or refute this hypothesis.

**CONCLUSION**

The enhancement of axial flow fan performance for air-cooled condenser (ACC) fans using tip vortex manipulators are evaluated. The study utilizes 3D RANS CFD simulation and experimental tests to perform the investigation. Analyses consider differences in performance when simulating and testing an unmodified and modified ACC scale fan. The modification of the fan entails the attachment of blade tip endplates that aim to manipulate the tip leakage flow structures to improve fan performance compared to the unmodified fan. The effects of two endplate designs in this regard are assessed.

Experimental results indicate that fan performance can be improved when incorporating blade tip modifications. Increases of up to 5.2% at design point and 2.3% for maximum fan static efficiency are achieved. Between the two endplate designs considered for the B2a fan, the "Adjusted Erasmus" endplate achieves greater performance enhancements than the "TLV stabilization" endplate.

Simulation results show good correlation with experimental results for the unmodified fan but fail to show enhancement of fan performance under modification.

An analysis of the numerical results reveals that tip vortical structures are manipulated by the endplates, while some observable effects on the tip leakage flows also agree with findings in literature. The evaluation however does not establish a clear reason for the inability of the simulation model to predict improvements in fan performance when modified.

A hypothesis is presented on the abovementioned shortcoming of the numerical model. The hypothesis notes deflection of the test fan blade and endplates during operation, which is not captured by the model, as potential reason for the discrepancy in results. Future numerical work should consider simulating the fan in its deflected state to achieve closer agreement with experimental results.

**ACKNOWLEDGMENT**

This research is supported by the Solar Thermal Energy Research Group (STERG) of Stellenbosch University.

**CONFLICT OF INTEREST**

The authors confirm that they are not aware of any conflicts of interest.

**DATA AVAILABILITY STATEMENT**

All simulated and experimental data presented in this paper may be obtained from the corresponding author upon reasonable request.

**NOMENCLATURE**

**Symbols**

| | | |
|---|---|---|
| $A$ | Area [ $m^2$ ] |
| $d$ | Diameter [ $m$ ] |
| $H$ | Helicity |
| $k$ | Turbulent kinetic energy [ $J/kg$ ] |
| $\dot{m}$ | Mass flow rate [ $kg/s$ ] |
| $N$ | Rotational speed [ $rpm$ ] |
| $P$ | Power [ $W$ ] |
| $p$ | Pressure [ $Pa$ ] |
| $R$ | Specific gas constant [ $J/kgK$ ] |

| | | |
|---|---|---|
| $r$ | | Radius [ $m$ ] |
| $T$ | | Temperature [ °C ] |
| $\dot{V}$ | | Volumetric flow rate [ $m^3/s$ ] |
| $u$ | | Absolute velocity [ $m/s$ ] |

**Greek symbols**

| | | |
|---|---|---|
| $\alpha\varepsilon$ | | Compound calibration factor |
| $\Delta$ | | Difference; change in |
| $\varepsilon$ | | Turbulent dissipation rate [ $m^2/s^3$ ] |
| $\eta$ | | Efficiency |
| $\rho$ | | Density [ $kg/m^3$ ] |
| $\tau$ | | Torque [ $Nm$ ] |
| $\phi$ | | Flow coefficient |
| $\Psi$ | | Pressure coefficient |

**Subscripts**

| | | |
|---|---|---|
| $a$ | | Air |
| $amb$ | | Ambient; atmospheric |
| $bell$ | | Bell mouth |
| $c$ | | Circumferential |

| | |
|---|---|
| $d$ | Dynamic |
| $F$ | Fan |
| $norm$ | Normalized |
| $plen$ | Plenum; settling chamber |
| $s$ | Static |
| $t$ | Tip |
| $tot$ | Total |
| $z$ | Axial direction |

**Superscripts**

| | |
|---|---|
| $'$ | Scaled/ adjusted |

**Figure Captions List**

## Table Caption List

Table 1        B2a fan specifications, adapted from [16]

Table 2        Measurement accuracies, adapted from [17]

Table 3        Specific boundary condition values of numerical model

Table 4        Element count and major quality metrics of selected mesh for the various simulation cases

Table 5        Solution method and spatial discretization settings in Ansys Fluent

Figure 1

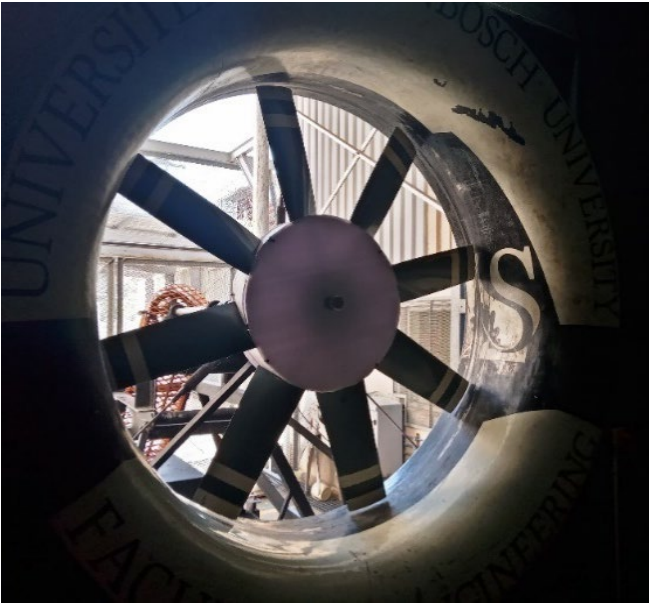

Figure 2

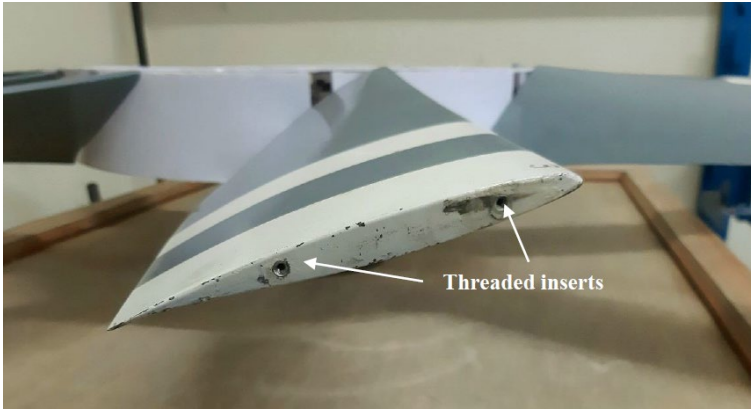

Figure 3

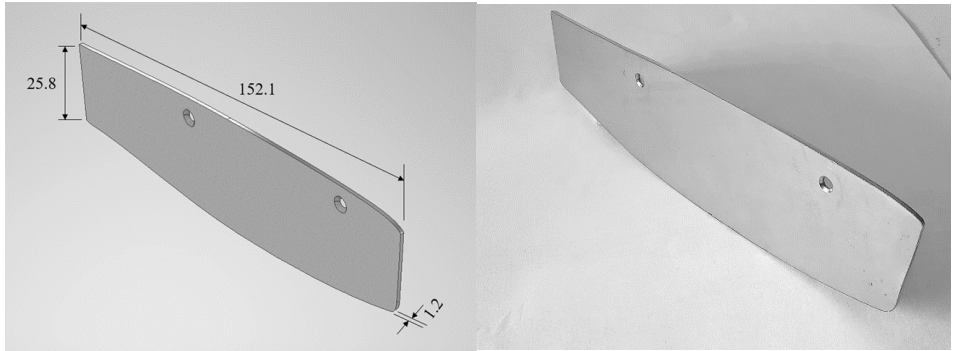

Figure 4

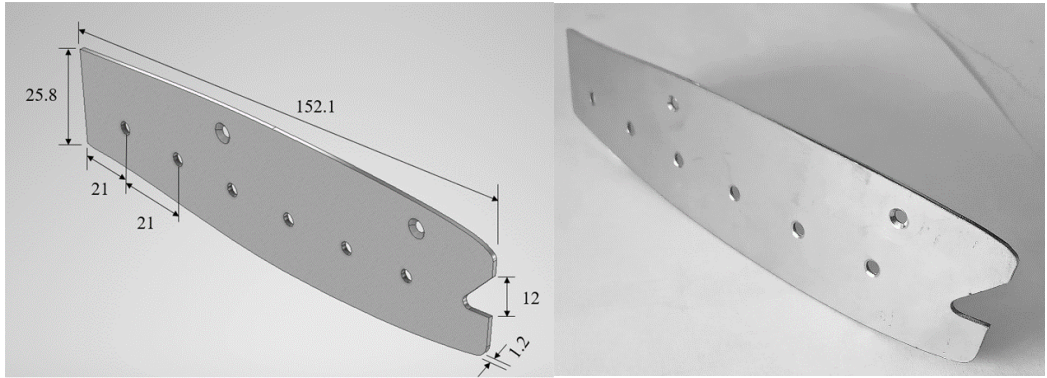

Figure 5

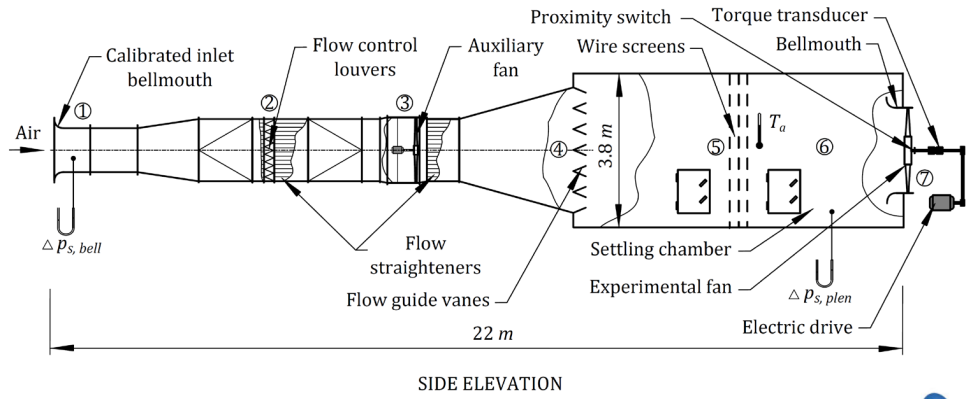

SIDE ELEVATION

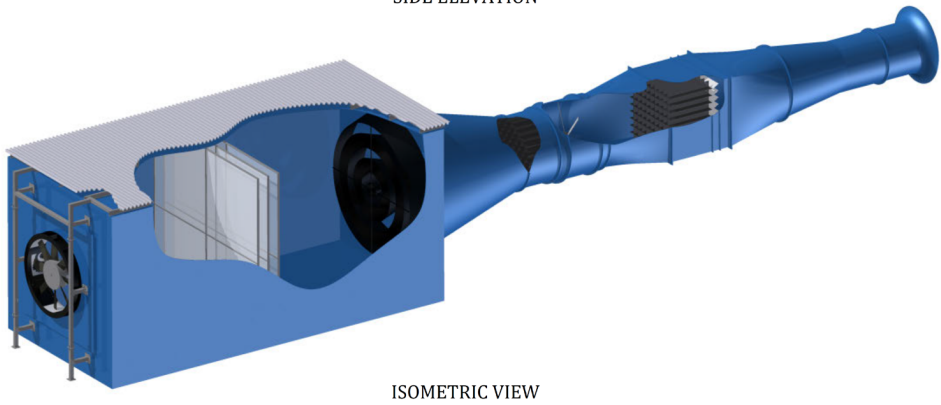

ISOMETRIC VIEW

Figure 6

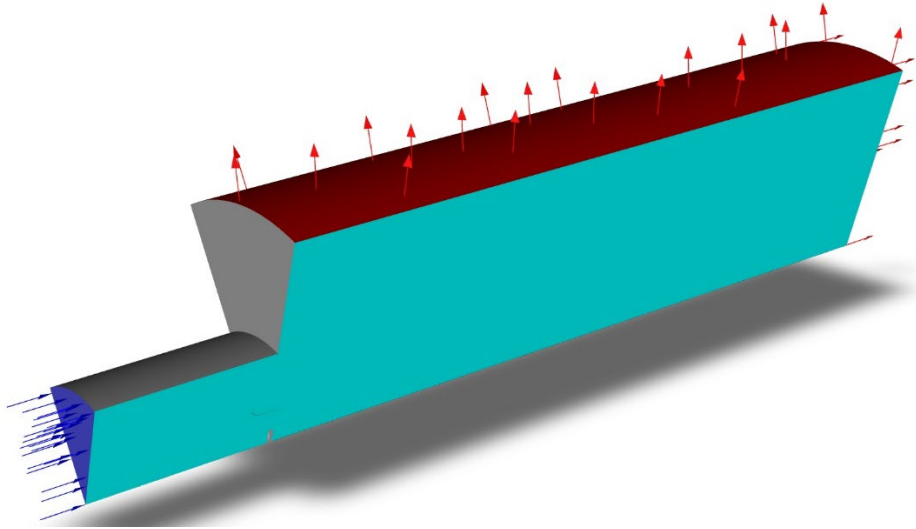

Figure 7

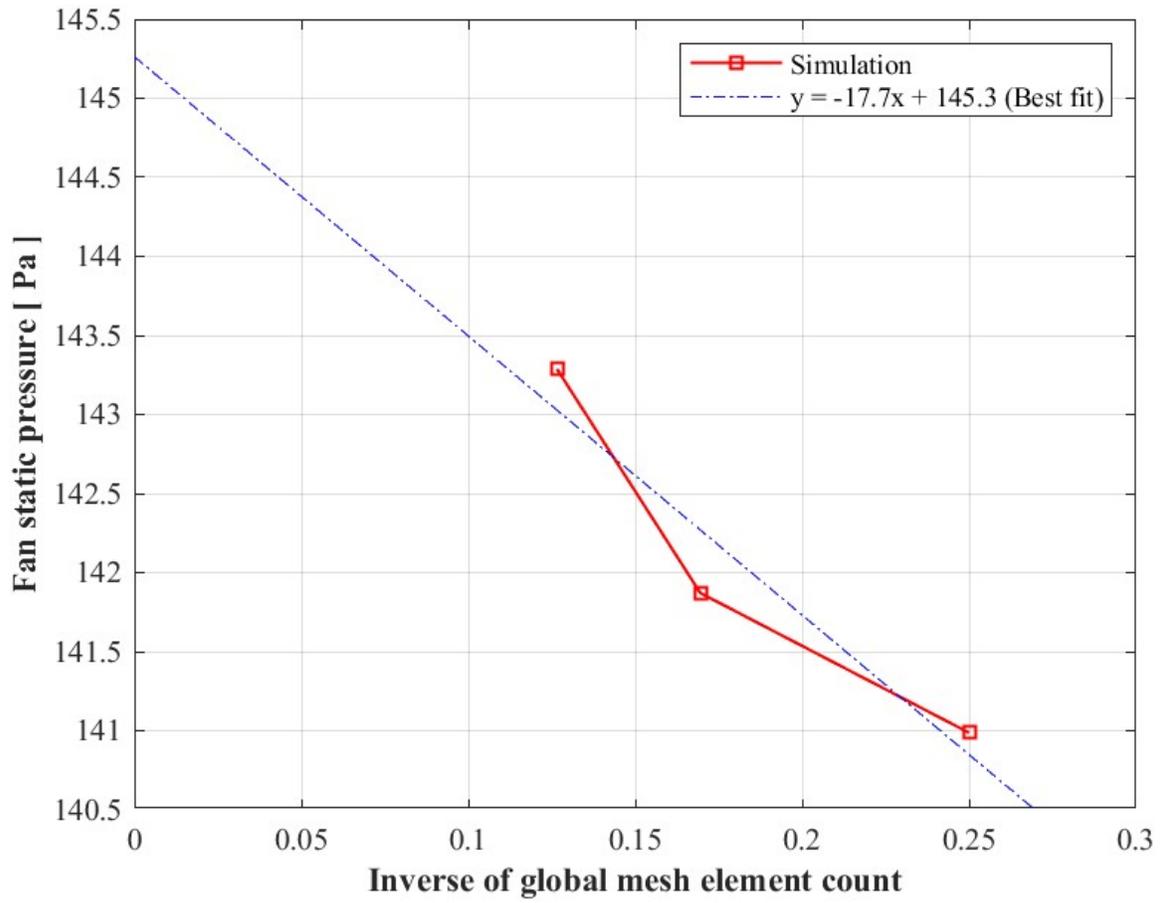

Figure 8

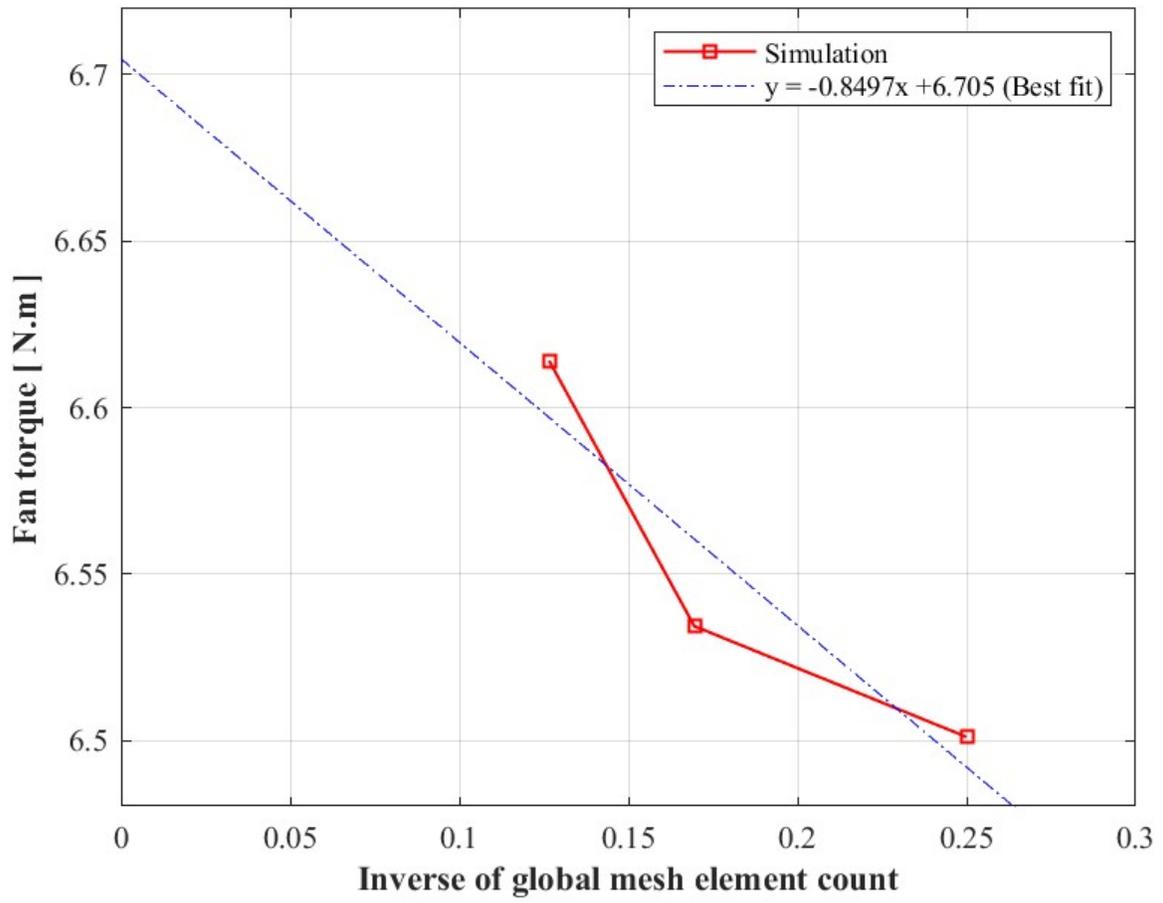

Figure 9

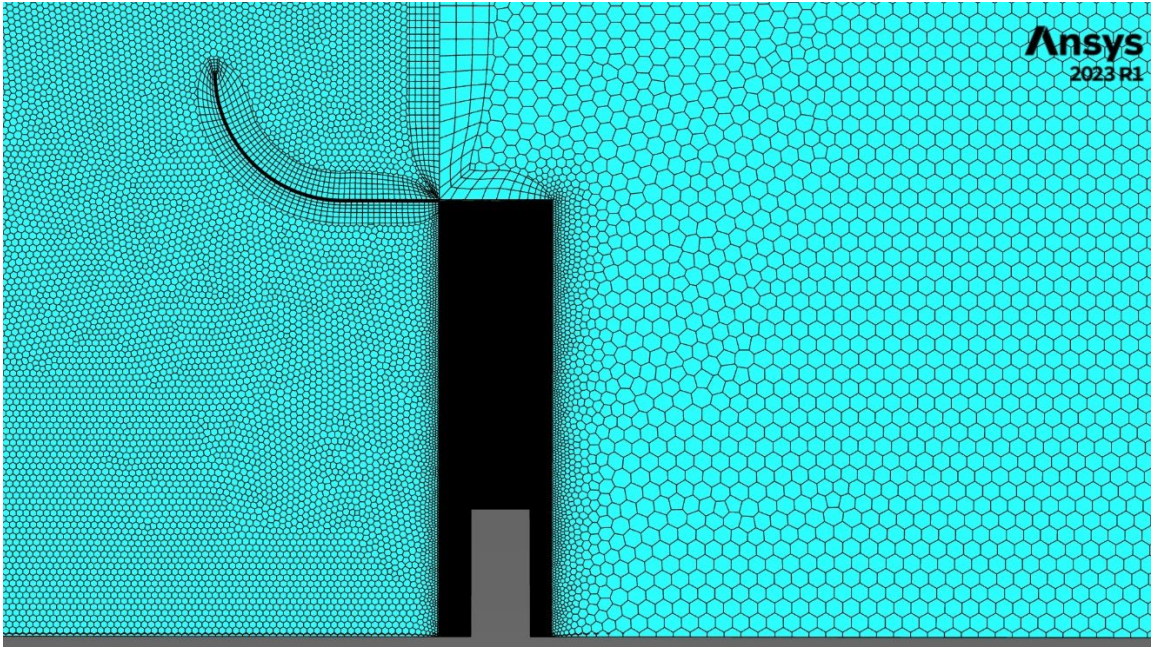

Figure 10

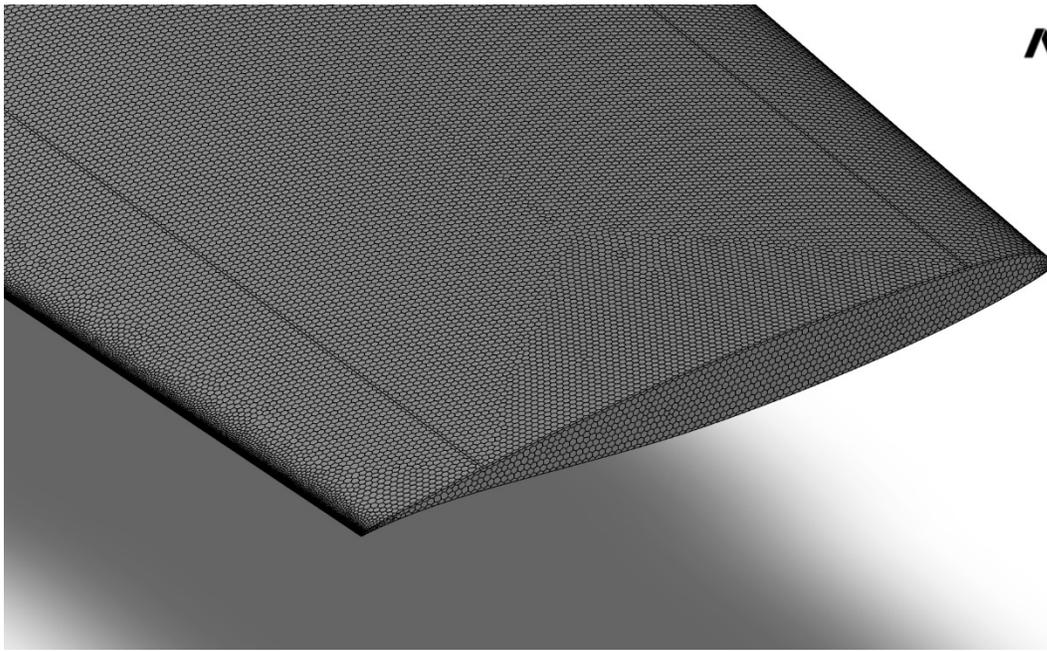

Figure 11

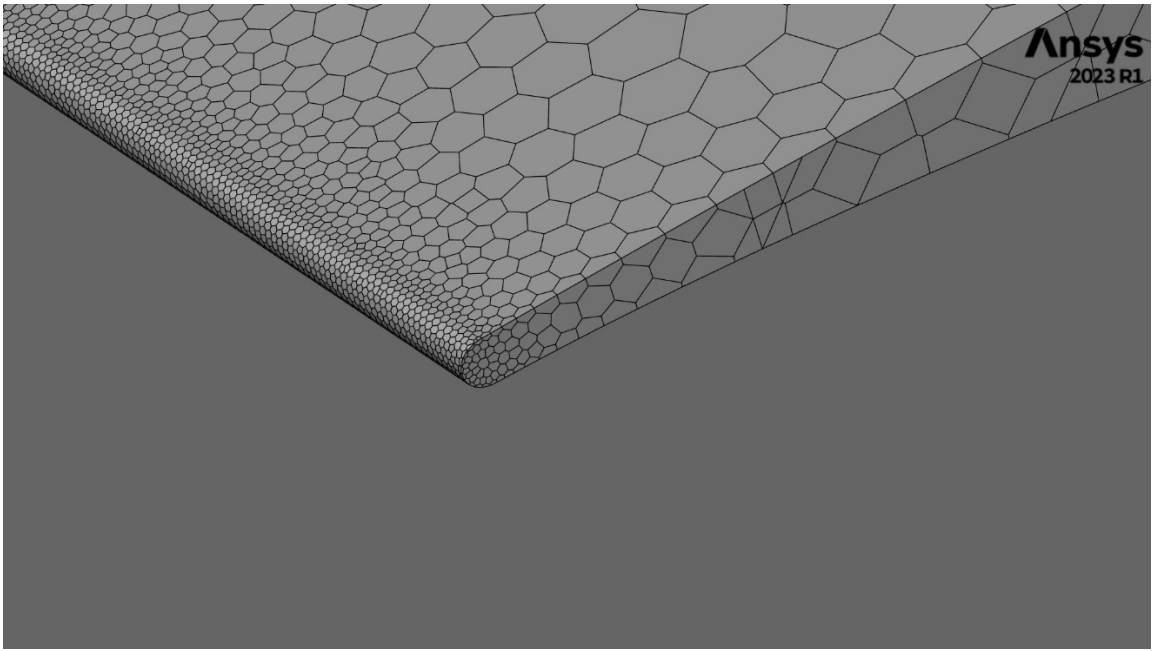

Figure 12

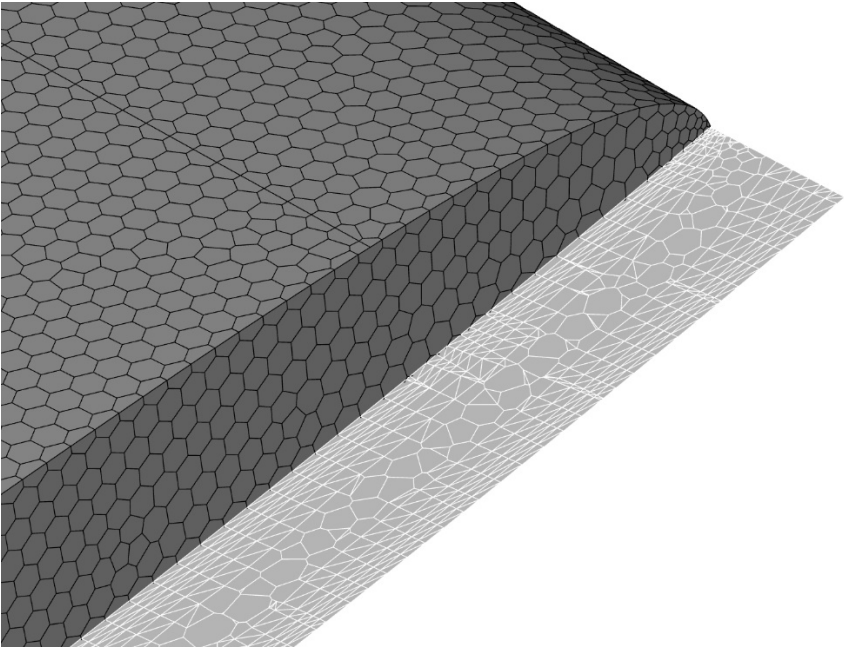

Figure 13

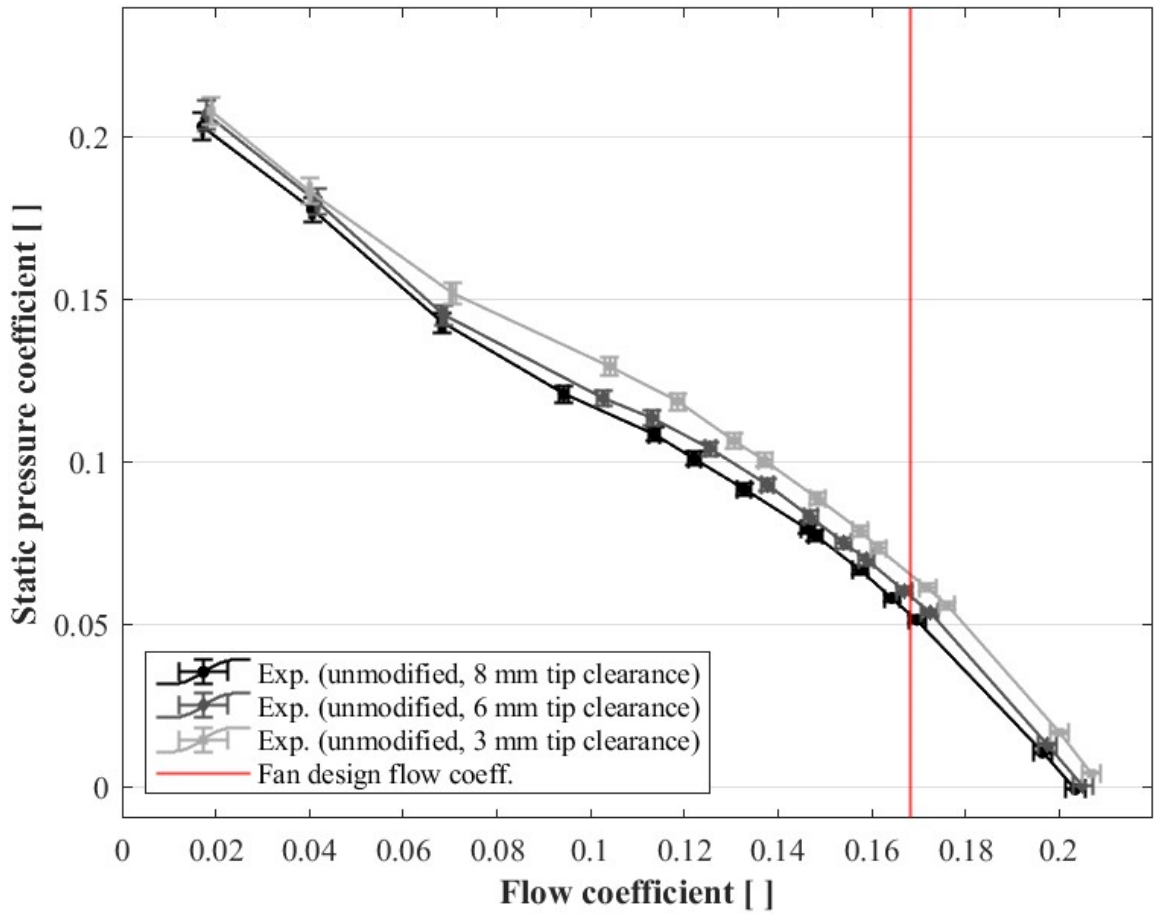

Figure 14

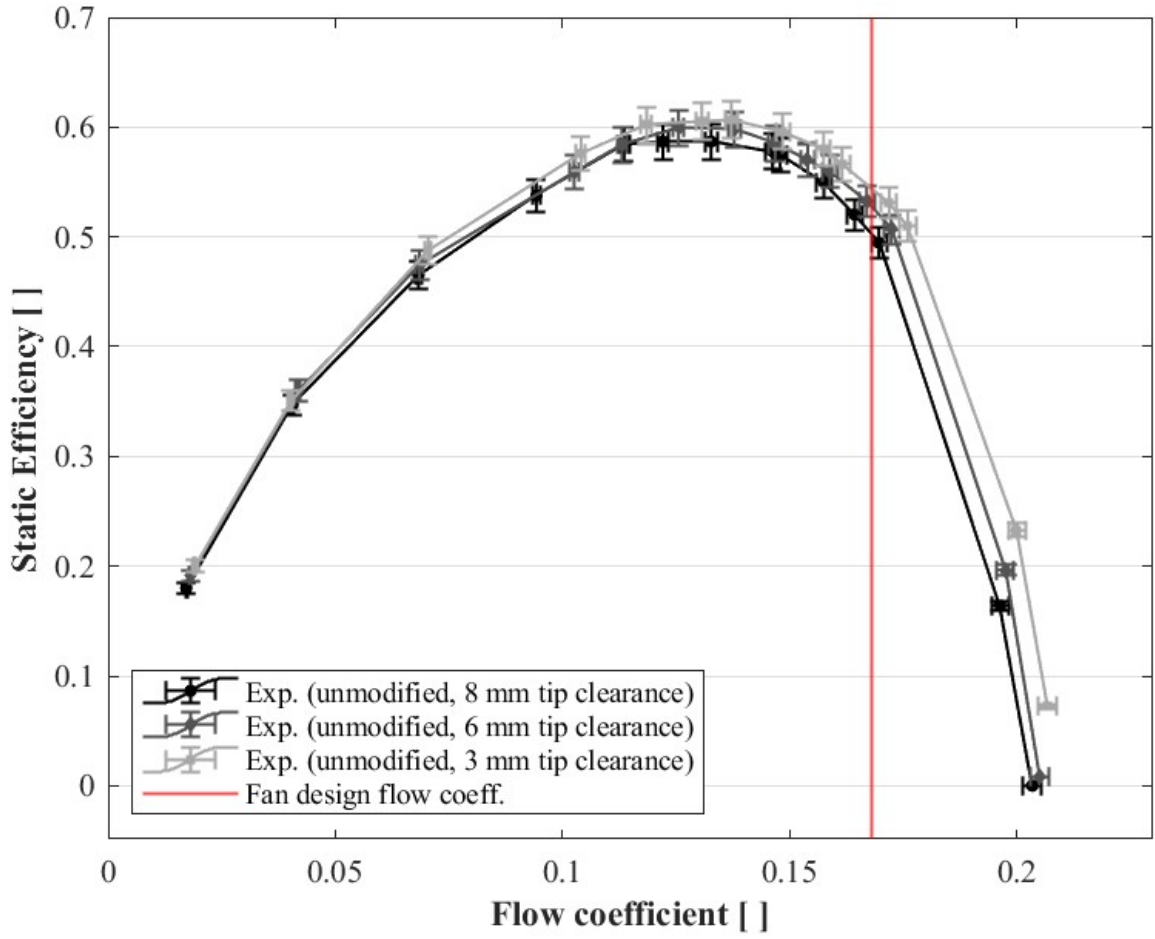

Figure 15

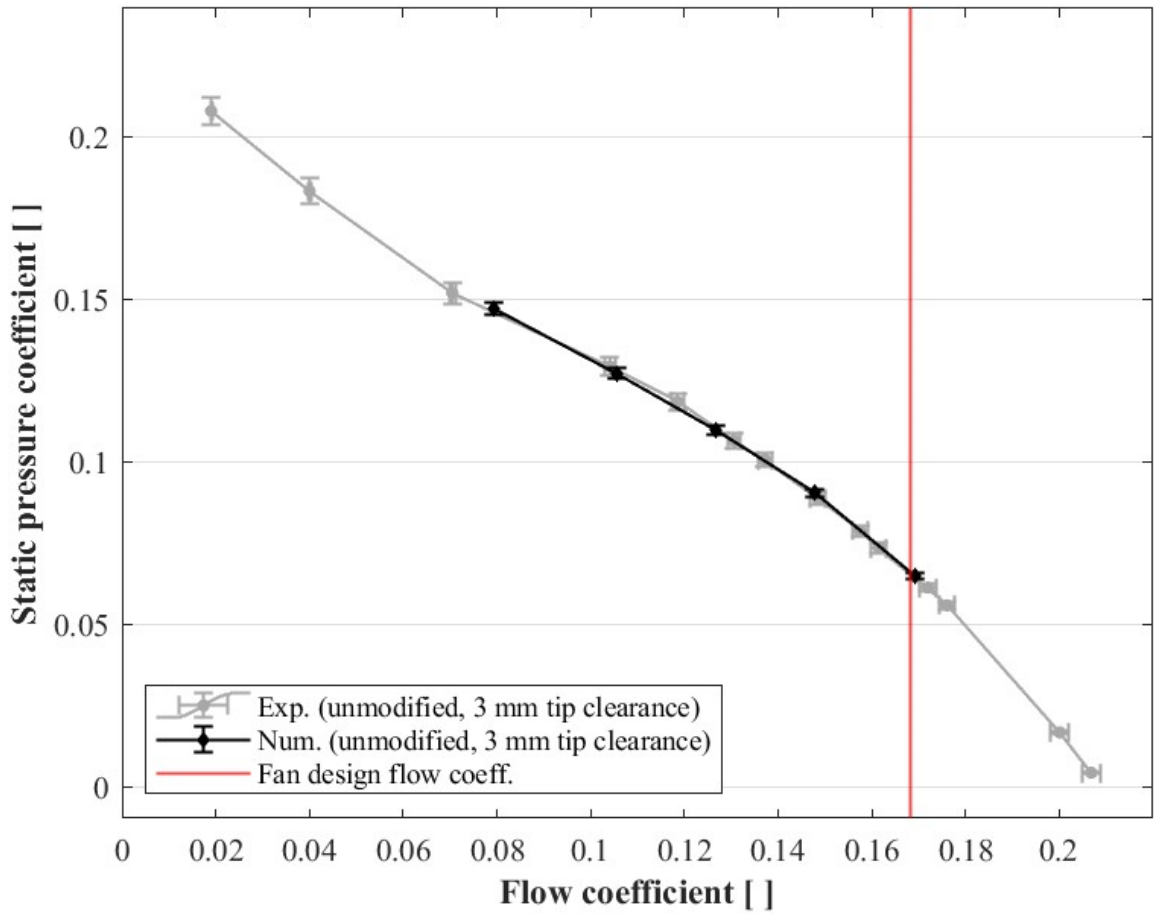

Figure 16

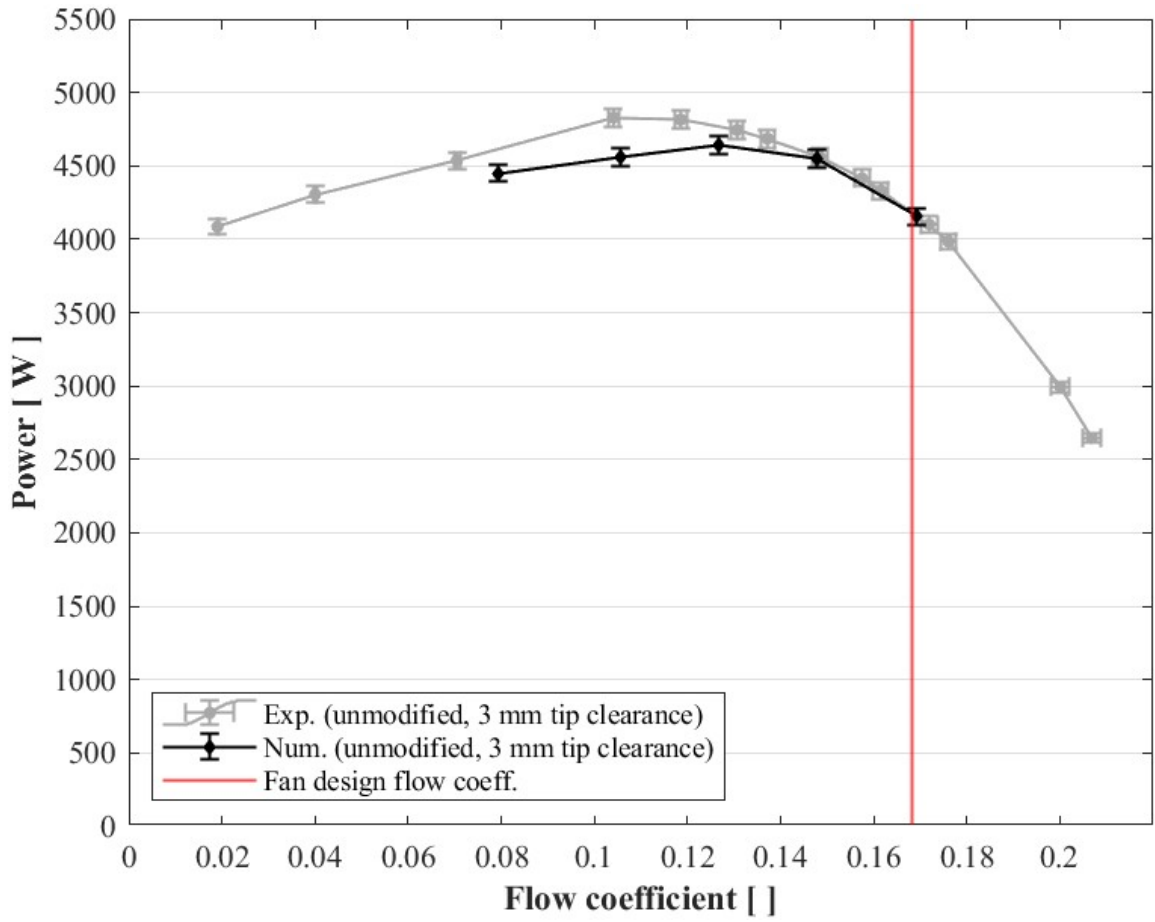

Figure 17

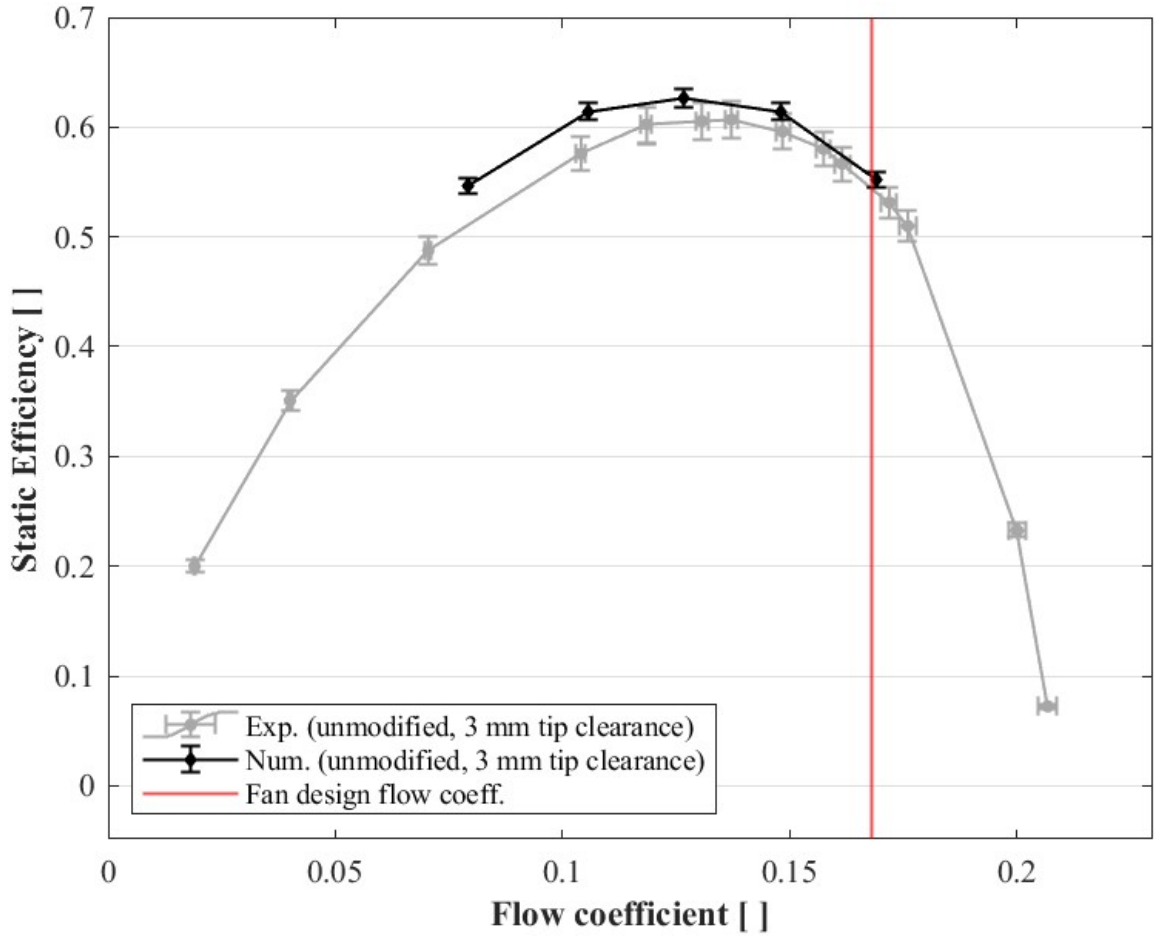

Figure 18

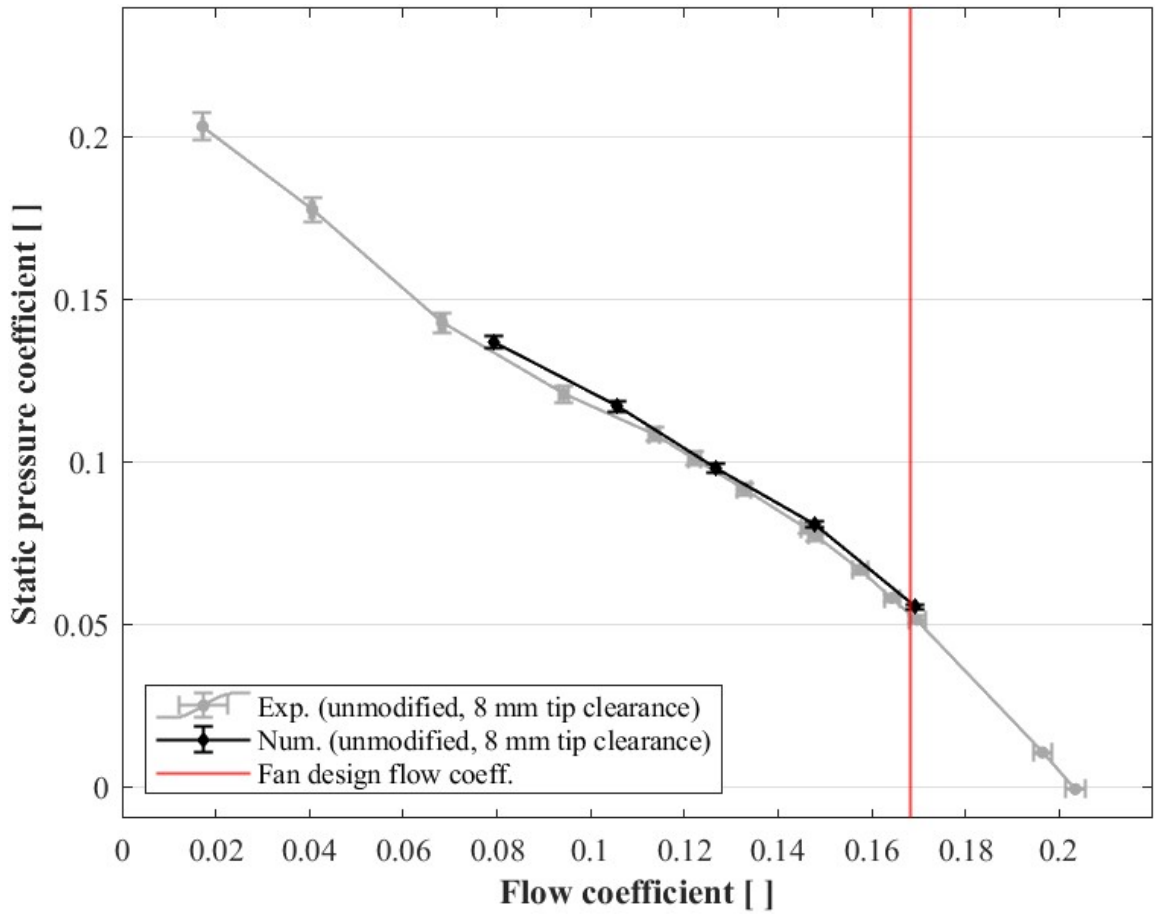

Figure 19

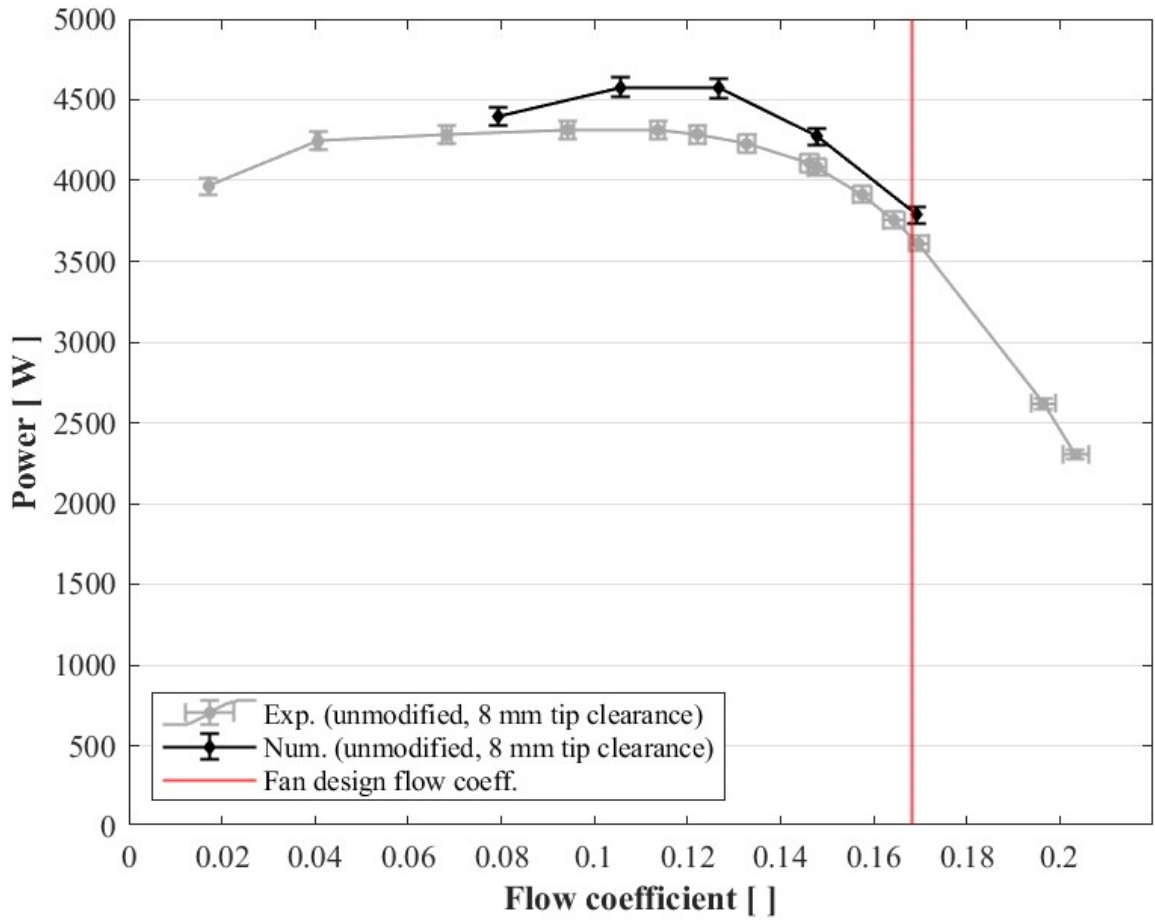

Figure 20

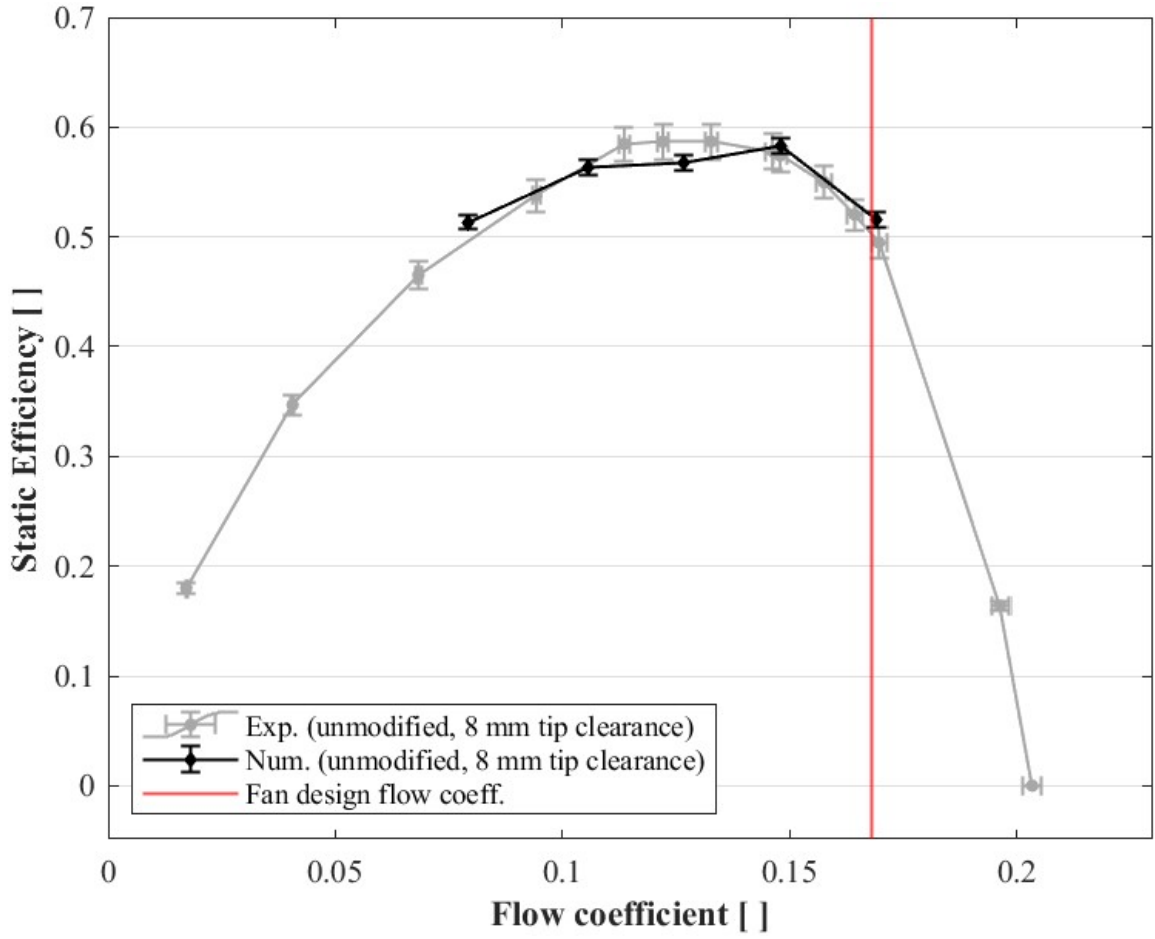

Figure 21

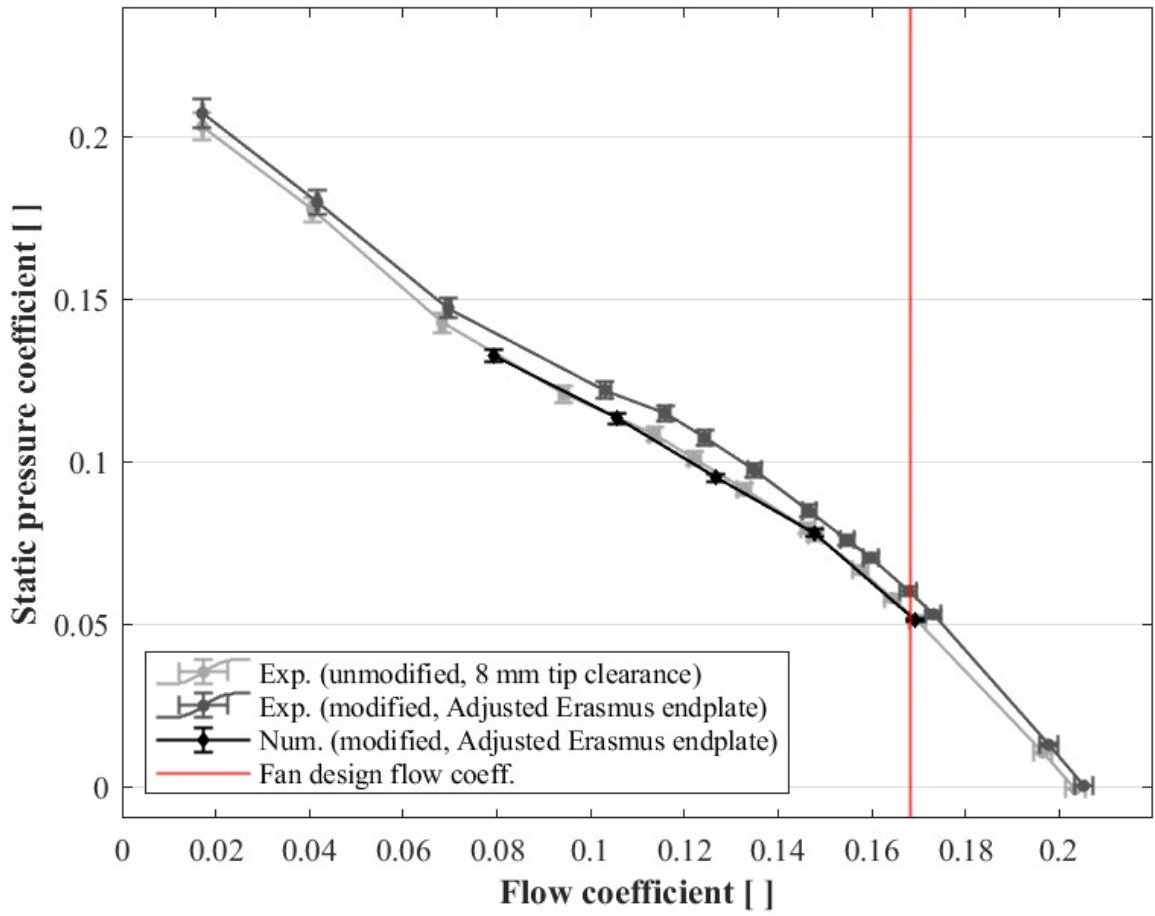

Figure 22

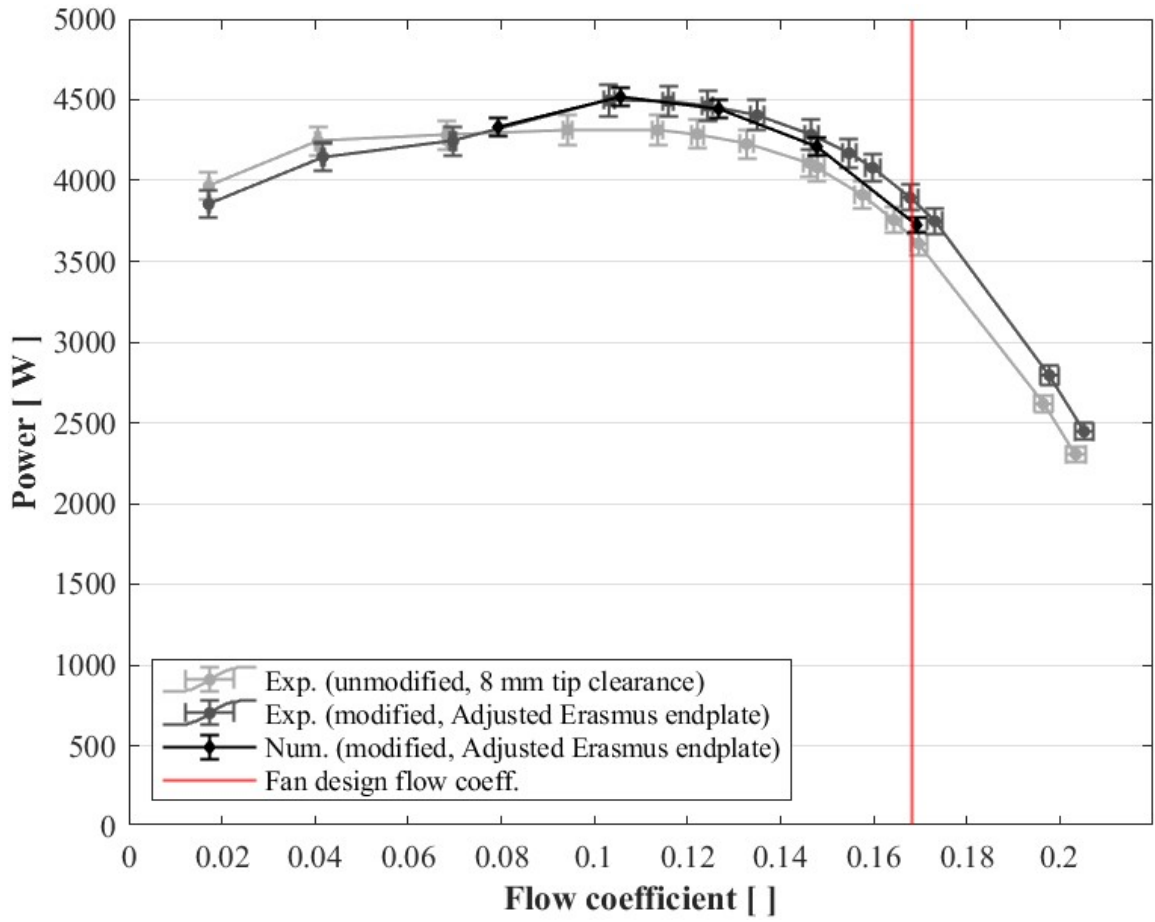

Figure 23

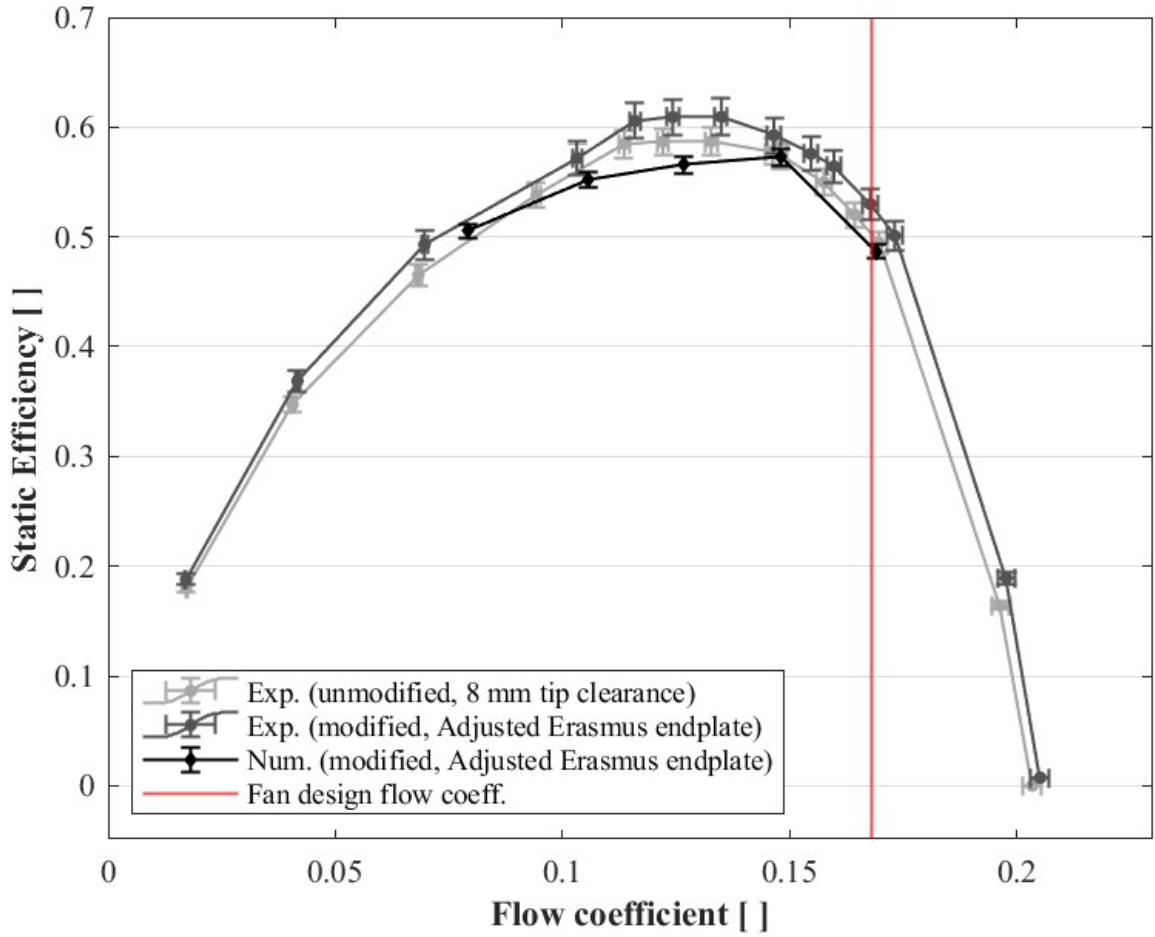

Figure 24

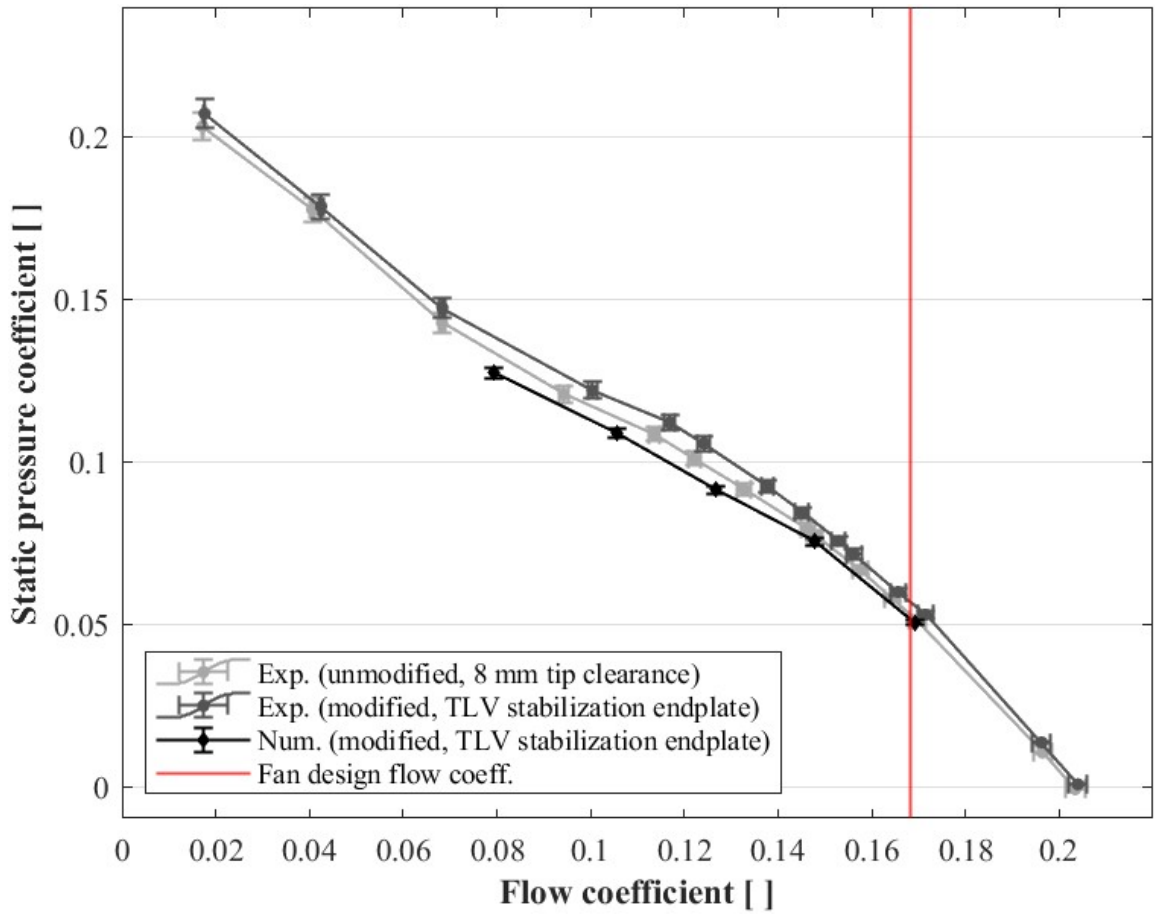

Figure 25

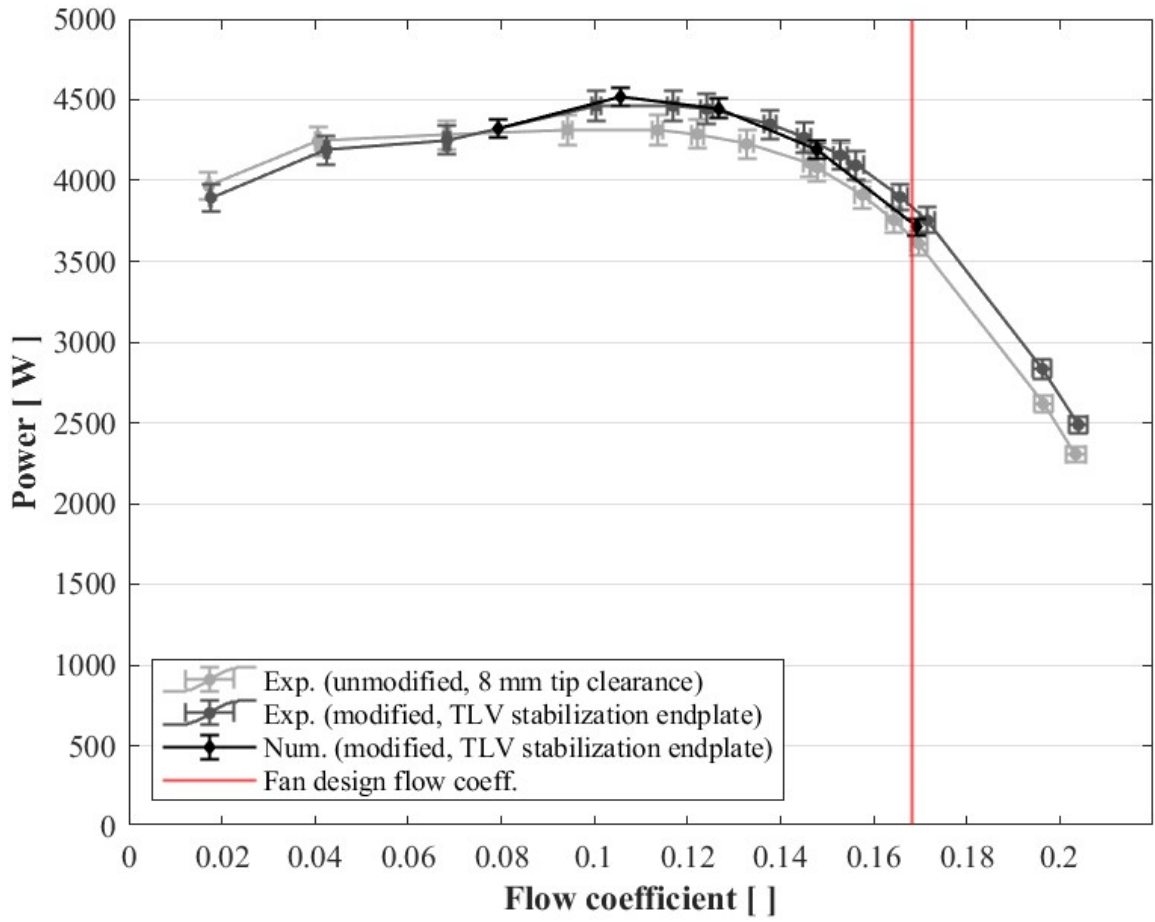

Figure 26

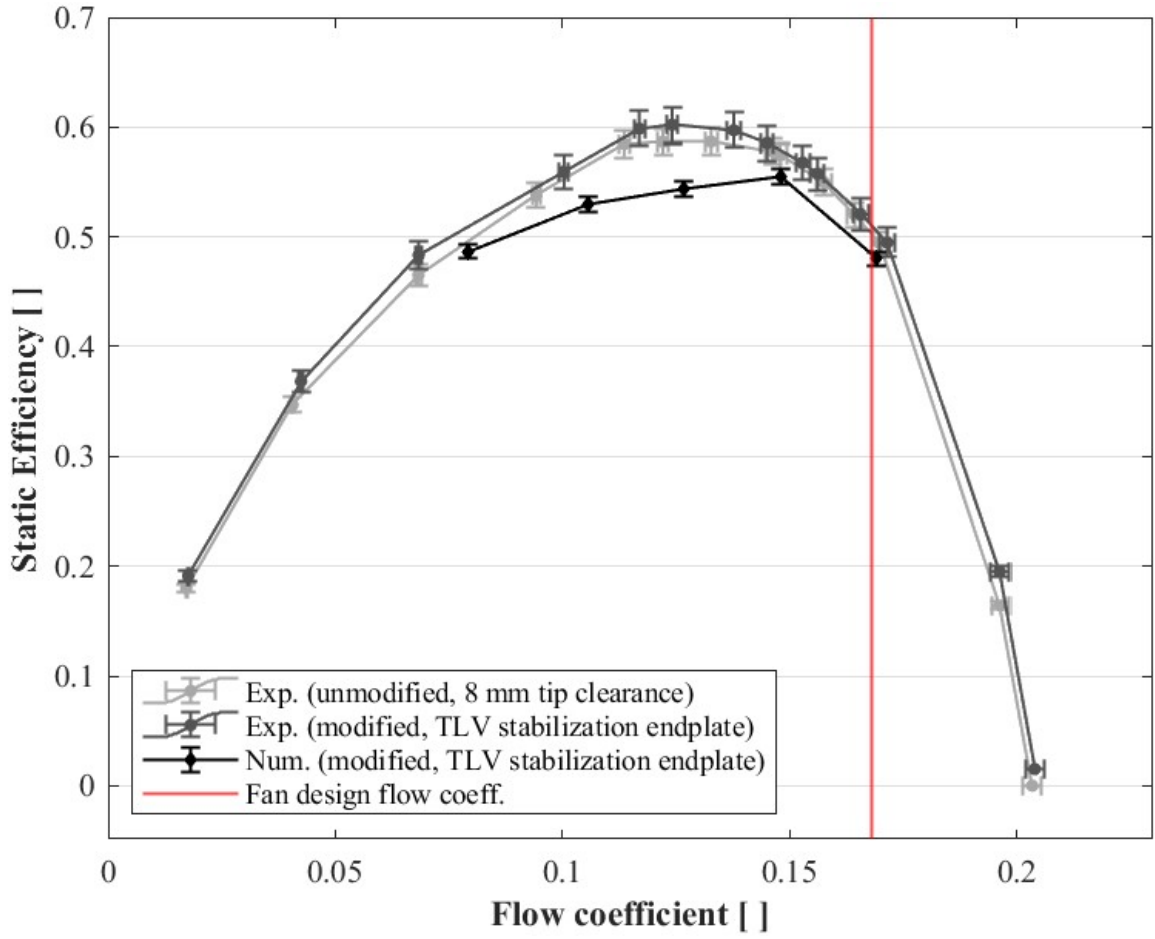

Figure 27

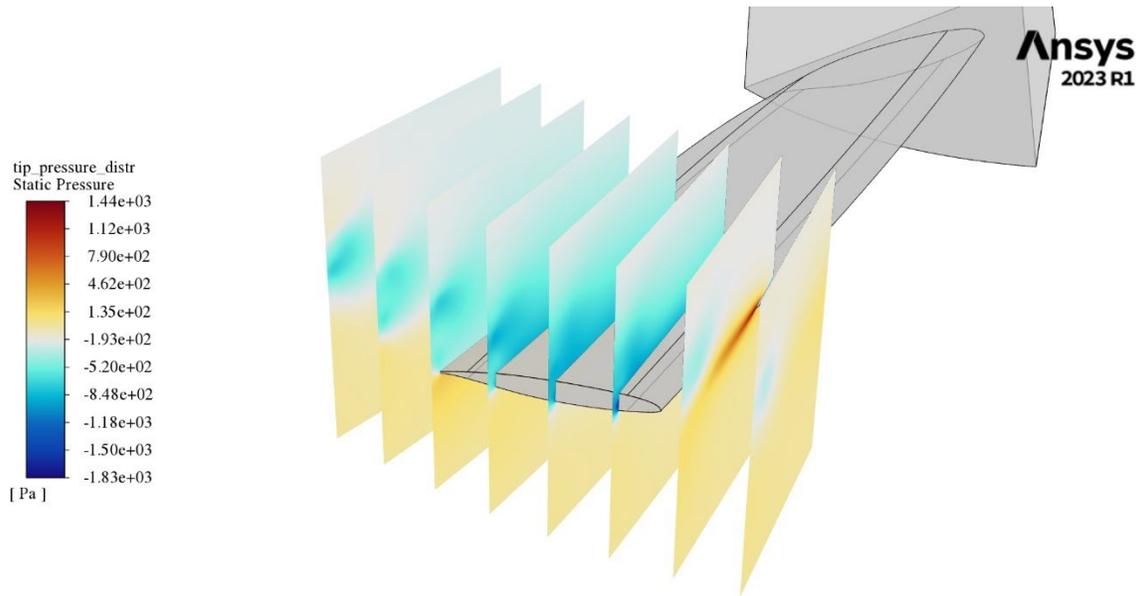

Figure 28

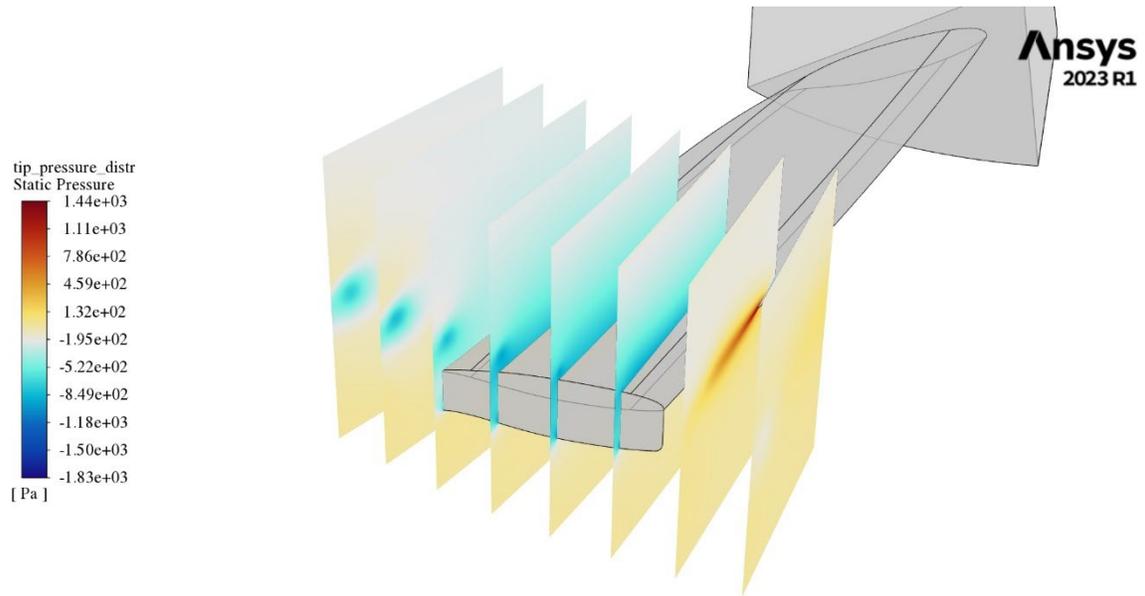

Figure 29

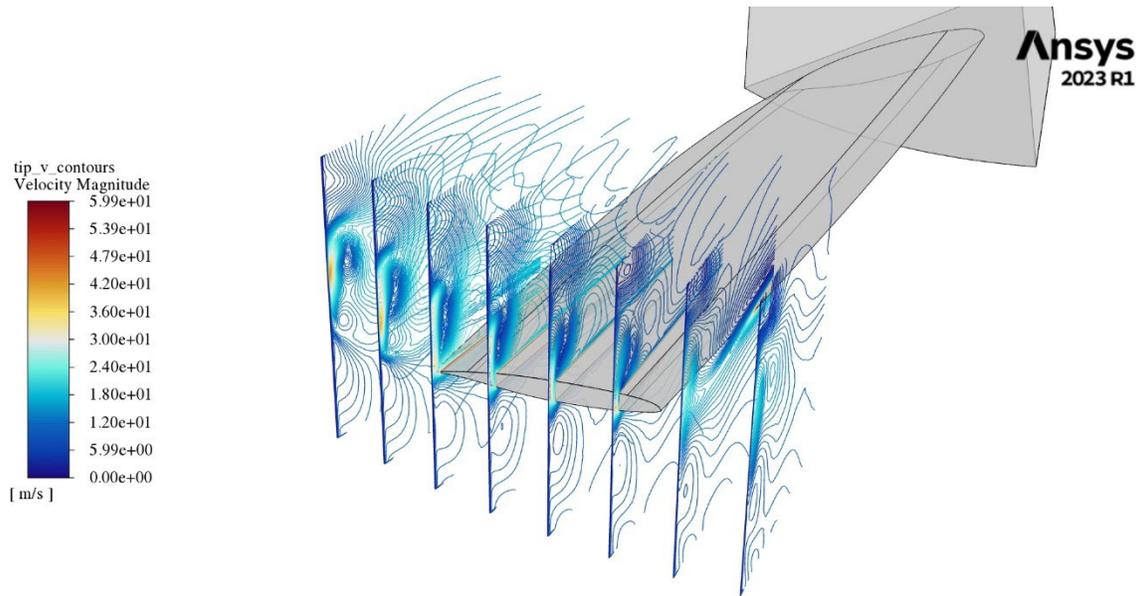

Figure 30

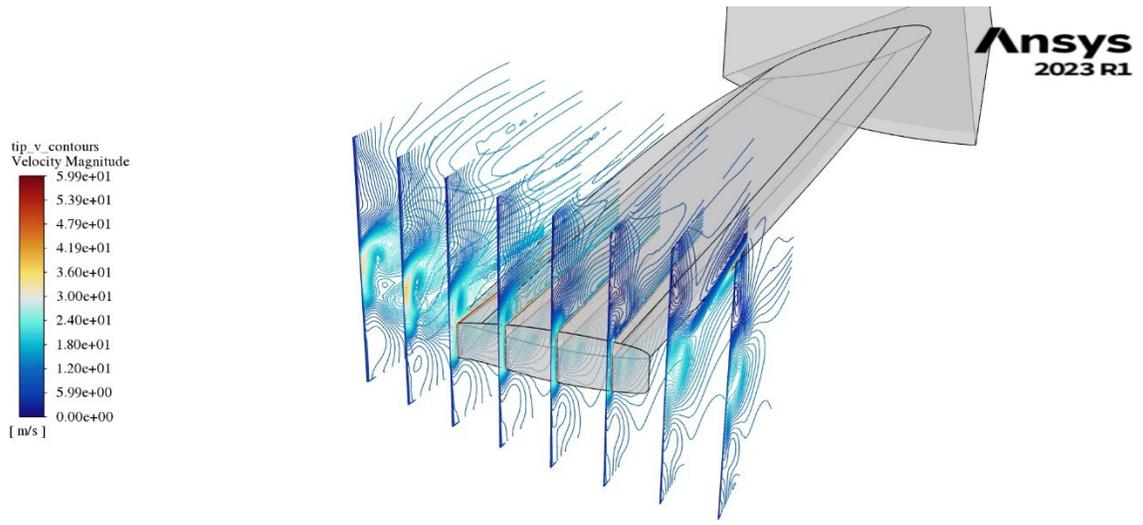

Figure 31

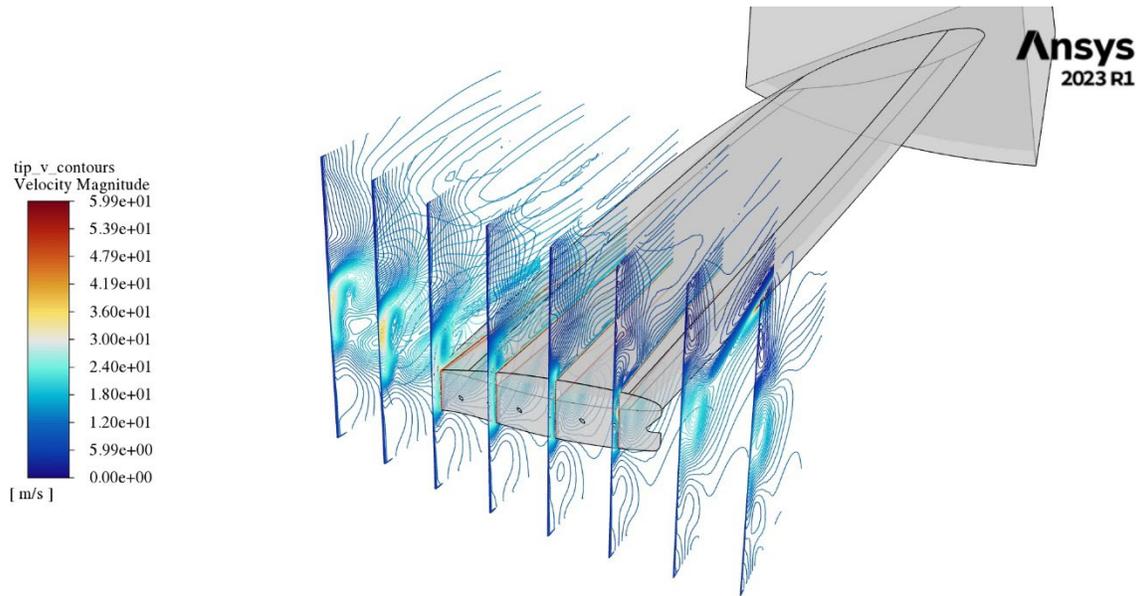

Figure 32

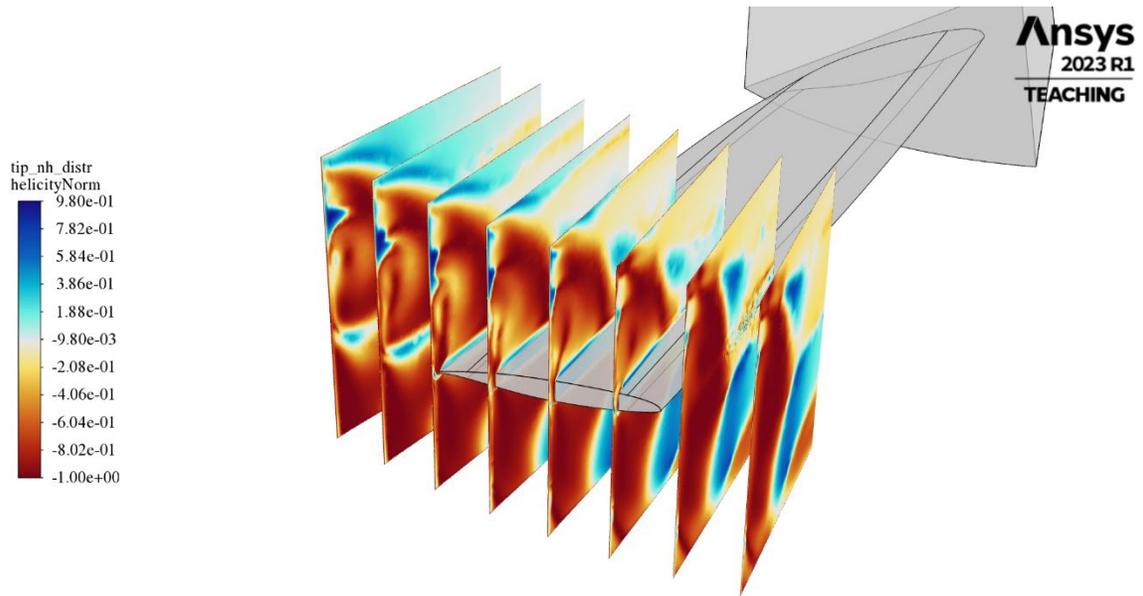

Figure 33

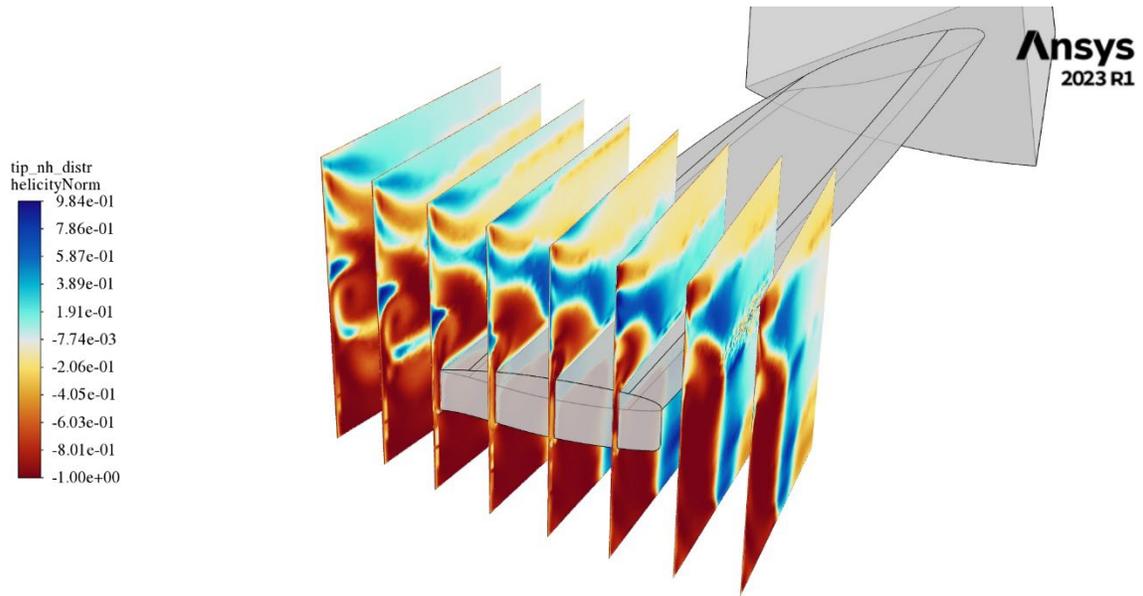

Figure 34

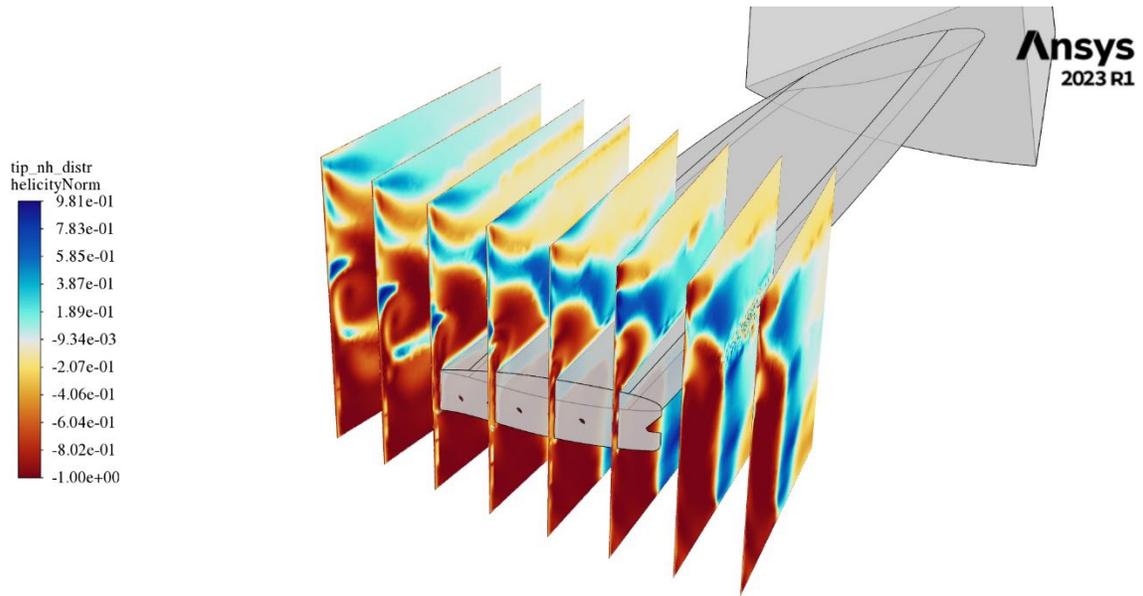

Table 1

| Design Parameter | Symbol | Value |
| --- | --- | --- |
| Casing diameter [ $m$ ] | | 1.542 |
| Number of blades | | 8 |
| Chord length at hub [ $m$ ] | | 0.184 |
| Chord length tip [ $m$ ] | | 0.153 |
| Hub-to-tip ratio | | 0.4 |
| Blade angle from hub chord, relative to tangential direction of blade rotation [ ° ] | | 31 |
| Hub diameter [ $m$ ] | | 0.6144 |
| Design tip clearance [ $mm$ ] | | 3 |
| Design static pressure rise [ $Pa$ ] | $\Delta p_{Fs}$ | 210 |
| Design static pressure coefficient | $\Psi_{Fs}$ | 0.095 |
| Design rotational speed [ $rpm$ ] | $N'$ | 750 |
| Design volumetric flow rate [ $m^3/s$ ] | $\dot{V}$ | 16 |
| Design flow coefficient | $\phi$ | 0.168 |
| Design air density [ $kg/m^3$ ] | $\rho'$ | 1.2 |

Table 2

| Measurement | Accuracy |
|---|---|
| Bell mouth static pressure difference ( $\Delta p_{s,bell}$ ) | ±1.9% |
| Settling chamber static pressure difference ( $\Delta p_{s,plen}$ ) | ±2.1% |
| Shaft torque ( $\tau$ ) | ±1.3% |
| Rotational speed ( $N$ ) | Not available |
| Ambient temperature ( $T_{amb}$ ) | ±2.4% |
| Ambient pressure ( $p_{amb}$ ) | ±0.03% |

Table 3

| Domain | Cell zone/ boundary condition | Value |
|---|---|---|
| Inlet | Mass flow rate | 2.4 $kg/s$ |
| | Turbulence intensity | 3%, 0.01 m length scale |
| Rotor | Rotational speed | 750 $rpm$ |
| Outlet | Static pressure (gauge) | 0 $Pa$ |
| | Turbulence intensity | 3%, 0.01 m length scale |

Table 4

| Parameter | Inlet domain | Unmodified fan rotor domain (8 mm tip clearance) | Modified fan rotor domain (Adjusted Erasmus) | Modified fan rotor domain (TLV stabilization) | Outlet domain |
|---|---|---|---|---|---|
| Element count | 1.47m | 4.51m | 4.87m | 4.83m | 1.85m |
| Min. orthogonal quality | 0.171 | 0.163 | 0.157 | 0.151 | 0.180 |
| Min. cell size [ $m$ ] | 0.01 | 5.8e-5 | 5.8e-5 | 5.8e-5 | 0.004 |
| Max. cell size [ $m$ ] | 0.012 | 0.0014 | 0.0014 | 0.0014 | 0.03 |

Table 5

| Setting | Scheme |
| --- | --- |
| Pressure-velocity coupling | Coupled |
| Gradient | Green-Gauss node-based |
| Pressure interpolation | PRESTO! |
| Momentum | Second-order upwind |
| Turbulent kinetic energy | Second-order upwind |
| Turbulent dissipation rate | Second-order upwind |